\documentclass[a4paper,11pt]{article}
  
\usepackage{jcappub} 

\usepackage{hyperref}
\usepackage[dvipsnames]{xcolor}
\usepackage{amsmath}
\usepackage{amssymb}
\usepackage{mathtools}
\usepackage{physics}
\usepackage{siunitx}
\usepackage{tabularray} 
\UseTblrLibrary{booktabs}

\usepackage{cleveref}

\newcommand{\Pp}{\mathcal{P}}
\newcommand{\Cc}{\mathcal{C}}
\newcommand{\chipi}{{\chi\pi}}
\newcommand{\kc}{k_\sigma}
\newcommand{\MPl}{M_\text{Pl}}

\newcommand{\et}[1]{{\color{green}[Eemeli: #1]}}

\def\be#1\ee{\begin{equation}#1\end{equation}}
\def\bl#1\el{\begin{align}#1\end{align}}
\def\ba#1\ea{\begin{align*}#1\end{align*}}

\newcommand{\ee}{\mathrm{e}}

\newcommand{\Mpl}{M_\mathrm{Pl}}

\newcommand{\ttheta}{\tilde{\theta}}

\definecolor{MONZA}{HTML}{CF000F}
\definecolor{DARKBLUE}{HTML}{00008b}
\definecolor{DARKMAGENTA}{HTML}{8b008b}

\def\beqa{\begin{eqnarray}}
\def\eeqa{\end{eqnarray}}

\def\dd{{\rm d}}

\newcommand{\beq}{\begin{eqnarray}} 
\newcommand{\eeq}{\end{eqnarray}}

\title{Stochastic Axion-like Curvaton: \it{ Non-Gaussianity and Primordial Black Holes Without Large Power Spectrum }}

\author[a,b]{Chao Chen,}
\author[c]{Anish Ghoshal,}
\author[d,e]{Gianmassimo Tasinato,} 
\author[f,g]{Eemeli Tomberg}

\affiliation[a]{School of Science, Jiangsu University of Science and Technology, Zhenjiang, 212100, China}
\affiliation[b]{Jockey Club Institute for Advanced Study, The Hong Kong University of Science and Technology, Hong Kong, China}
\affiliation[c]{Institute of Theoretical Physics, Faculty of Physics, \\ University of Warsaw, ul. Pasteura 5, 02-093 Warsaw, Poland}
\affiliation[d]{Physics Department, Swansea University, SA28PP, United Kingdom}
\affiliation[e]{Dipartimento di Fisica e Astronomia, Universit\`a di Bologna, and \\
INFN, Sezione di Bologna, I.S.~FLAG, viale B.~Pichat 6/2, 40127 Bologna, Italy}
\affiliation[f]{Laboratory of High Energy and Computational Physics, NICPB, \\
R{\"a}vala pst. 10, Tallinn, 10143, Estonia}
\affiliation[g]{Consortium for Fundamental Physics, Physics Department, Lancaster University, \\ Lancaster LA1 4YB, United Kingdom}

\emailAdd{chaochen012@gmail.com}
\emailAdd{anish.ghoshal@fuw.edu.pl}
\emailAdd{g.tasinato2208.at.gmail.com}
\emailAdd{e.tomberg@lancaster.ac.uk}

\abstract{ 
We discuss a mechanism of primordial black hole (PBH) formation that does not require specific features in the inflationary potential, revisiting  previous literature.
In this mechanism, a light spectator field evolves stochastically during inflation and remains subdominant during the post-inflationary era. Even though the curvature power spectrum stays small at all scales, rare perturbations of the field probe a local maximum in its potential, leading to non-Gaussian tails in the  distribution of curvature fluctuations, and to copious PBH production. For a concrete axion-like particle (ALP) scenario we analytically determine the distribution of the compaction function for perturbations, showing that it is characterized by a heavy tail, which produces an extended PBH mass distribution. We find the ALP mass and decay constant to be correlated with the PBH mass, for instance, an ALP with a mass $m_a = \SI{5.4e14}{eV}$ and a decay constant $f_a = \num{4.6e-5}\Mpl$ can lead to PBHs of mass $M_\text{PBH} = \SI{e21}{g}$ as the entire dark matter (DM) of the universe, and is testable in future PBH observations via lensing in the NGRST and mergers detectable in the LISA and ET gravitational wave detectors. We then extend our analysis to mixed ALP and PBH dark matter and Higgs-like spectator fields. 
We find that  PBHs cluster strongly over all cosmological scales, clashing with CMB isocurvature bounds. We argue  that this problem is shared by all PBH production from inflationary models that depend solely on large non-Gaussianity without a peak in the curvature power spectrum and discuss possible remedies.}
\begin{document}
\maketitle
\flushbottom

\section{Introduction}

The phenomenon of cosmic inflation predicts the existence of primordial curvature perturbations, explaining the origin of the temperature fluctuations in the cosmic microwave background radiation (CMBR) and the large-scale structure (LSS) formation in the Universe. 
We know that the primordial curvature perturbation $\zeta$ is already present when the large cosmological scales enter the Hubble horizon in the early Universe.  But when it is outside the Hubble horizon, the Fourier components $\zeta_k$ ($k$ being the wavenumber) become time-independent, setting initial conditions for the CMB and LSS formation in the Universe. Usually the vacuum fluctuations of a scalar (or vector~\cite{Dimopoulos:2009vu}) field are generated $\zeta_k$ when they exit the horizon during cosmic inflation (that is, when $ k=aH\equiv \dot a$ where $a(t)$ is the scale factor of the FLRW metric). This scalar field can either be the inflaton itself, as is the case for the single-field inflation, or it can be a spectator field, like in the case of the well-known curvaton scenario~\cite{Lyth:2001nq,Moroi:2001ct,Enqvist:2001zp,Lyth:2002my,Sloth:2002xn,Dimopoulos:2003ii}. In the latter, the curvaton's contribution to the energy budget of the Universe is usually negligible during inflation, and $\zeta$ is generated only after inflation, when the curvatonic component becomes important.



From the observations of the CMB scale~\cite{Planck:2018vyg}, the amplitude of typical curvature perturbations is of the order $10^{-5}$ compared to the homogeneous background, 
while larger curvature perturbations may exist at smaller scales and may lead to high-density regions that collapse to primordial black holes (PBHs), see e.g., \cite{Hawking:1971ei,Carr:1974nx,Carr:1975qj,Ivanov:1994pa, GarciaBellido:1996qt, Kawasaki:1997ju, Yokoyama:1998pt, Khlopov:2008qy,Sasaki:2018dmp, Garcia-Bellido:2017mdw, Hertzberg:2017dkh, Cai:2018tuh, Cai:2019bmk, Fu:2019ttf, Liu:2021svg,Cai:2023uhc, Escriva:2022duf, Ge:2023rrq, Fu:2022ssq, Wang:2024vfv}.
PBHs may contribute to supermassive black holes~\cite{LyndenBell:1969yx, Kormendy:1995er} and the stellar-mass black hole merger events observed by gravitational wave (GW) detectors~\cite{LIGOScientific:2016dsl, LIGOScientific:2021djp}.
The formation of stellar mass PBHs may induce GW signatures that fall within the reach of current Pulsar Timing Array (PTA) experiments (see e.g.,~\cite{Cai:2023dls,Inomata:2023zup,Chang:2023vjk, Yi:2023mbm,Ellis:2023oxs, Wang:2023sij, Choudhury:2024dzw,Domenech:2024rks,Wang:2023ost}), in the nano-Hz frequency range, where PTA experiments (NANOGrav~\cite{NANOGrav:2020bcs}, CPTA~\cite{Xu:2023wog}, PPTA~\cite{Reardon:2023gzh}, EPTA~\cite{EPTA:2023fyk}, etc.) recently reported evidence for a stochastic common process.
At larger frequencies, LISA \cite{LISACosmologyWorkingGroup:2022jok} would be able to observe the imprints of asteroid mass PBH formation (see e.g., \cite{Saito:2008jc,Garcia-Bellido:2017aan,Cai:2018dig,Bartolo:2018evs}). These PBHs, with masses of $10^{17\text{--}23} $g, are a favorable candidate for dark matter (DM) \cite{Chapline:1975ojl, Carr:2016drx,Inomata:2017okj,Inomata:2017vxo}.

In single-field inflation, large curvature perturbations are generated when the inflaton rolls through a flat Section of its potential in so-called ultra-slow-roll inflation, see e.g.,~\cite{Dimopoulos:2017ged,Karam:2022nym,Ozsoy:2023ryl,Ghoshal:2023wri}. These scenarios typically involve a lot of fine-tuning, as discussed in detail in \cite{Cole:2023wyx}, first to obtain a peak in the curvature power spectrum of order $10^{-2}$ and then to adjust the peak for exactly the desired PBH abundance. In recent years, it has been shown that curvaton models\footnote{Also dynamics of spectator fields like the waterfall during hybrid inflation may lead to significant PBH formation Refs. \cite{Afzal:2024xci,Afzal:2024hwj}. } can also produce large perturbations responsible for PBH generation \cite{Kohri:2012yw,Kawasaki:2012wr,Kawasaki:2013xsa,Bugaev:2013vba,Ando:2017veq,Ando:2018nge,Chen:2019zza,Cai:2021wzd,Inomata:2020xad,Liu:2020zzv,Zhou:2020kkf,Pi:2021dft,Liu:2021rgq,Kawasaki:2021ycf,Chen:2023lou,Ferrante:2023bgz,Hooper:2023nnl}.

In this paper, we will investigate a mechanism involving a light curvaton field or spectator field in general. The spectator field or the curvaton field may have a variety of microscopic origins and field theoretical properties since they  generic in various high-energy physics frameworks such as supersymmetry (SUSY), supergravity (SUGRA), grand unified theories (GUT), string theory, extra-dimensional models, etc \cite{Martin:2013tda}. Because the spectator field is light, the exact shape of its potential is somewhat irrelevant for its inflationary evolution, during which the field fluctuations are dominated by its quantum fluctuations. This evolution we will encapture in non-perturbative framework by a stochastic computation.
However, the potential becomes relevant after inflation, when the field starts to evolve classically. Eventually, the curvaton decays, transferring its fluctuations into curvature perturbations. In our setup, the typical curvature perturbations produced this way are small, below the observed CMB amplitude, which we assume to originate from other fields. However, in rare regions of space, the spectator's inflationary motion leaves it in a flat section of its potential, leading to a significant change in the local expansion history before decay and a strong curvature fluctuation. PBHs are produced later, when these fluctuations re-enter inside the Hubble radius and collapse gravitationally. 

The observed primordial perturbations follow a Gaussian distribution to a high accuracy \cite{Planck:2019kim}. However, PBHs form from strong, rare fluctuations far in the tail of the distribution, where non-Gaussianities may be important, see e.g., \cite{Young:2013oia, Franciolini:2018vbk, Cai:2022erk}. Some non-Gaussianity arises from the transformation between the curvature perturbations and the PBH-forming density perturbations, see e.g. \cite{Harada:2015yda,DeLuca:2019qsy,Young:2019yug}; some non-Gaussianity may be already present in the primordial curvature perturbations. Primordial non-Gaussianity is also present in the curvaton scenario \cite{Bartolo:2003jx,Bartolo:2005fp,Sasaki:2006kq,Enqvist:2008gk,Kohri:2009ac,Chingangbam:2009xi,Huang:2010cy,Kawasaki:2011pd,Fonseca:2011aa,Kawasaki:2012gg,Enomoto:2012uy,Mukaida:2014wma,Liu:2020zlr,Fong:2023egk,Ghoshal:2023lly}, and it is usually treated perturbatively, described by the local non-Gaussianity parameter $f_\text{NL}$ and its higher-order counterparts. They describe small deviations from Gaussianity. As mentioned above, we do not have a peak in the curvature power spectrum, so the PBH-forming perturbations must be far from Gaussian. For this reason, we solve the non-Gaussianities non-perturbatively, using the $\delta\mathcal{N}$ formalism \cite{Sasaki:1995aw, Sasaki:1998ug, Wands:2000dp, Lyth:2004gb}, without resorting to an $f_\text{NL}$ expansion.


In~\cite{Carr:2019hud, Stamou:2023vft, Stamou:2023vwz, Stamou:2024xkk}, the authors considered a similar PBH production mechanism. However in their setup, the curvaton dominates over radiation in the PBH-forming patches, whereas we will see that even a sub-dominant curvaton may be enough for PBH formation. We obtain an analytical formula for the PBH density, depending on a few key parameters of our example model. 
We also show how phase transitions can introduce a natural lower cutoff in the PBH mass spectrum, and the clustering of PBHs formed through such mechanisms is also discussed.

As a specific example, we study an axion-like particle (ALP) as the curvaton \cite{Kawasaki:2012wr,Ando:2017veq,Inomata:2020xad,Ghoshal:2023lly}, with a sinusoidal potential whose maxima can produce PBHs through the above-described mechanism. The axion-like field is the angular component $\vartheta$ of a generic complex scalar field $\Phi=\varphi e^{i \vartheta}$ charged under a $U(1)_{PQ}$ symmetry. Historically, the so-called Peccei-Quinn (PQ) mechanism~\cite{Peccei:1977hh,Peccei:1977ur} (for reviews, see \cite{Jaeckel:2010ni,Ringwald:2012hr,Arias:2012az,Graham:2015ouw,Marsh:2015xka,Irastorza:2018dyq, DiLuzio:2020wdo}) was developed to address the strong CP problem of quantum chromodynamics (QCD). It predicts the existence of a light pseudo-Nambu-Goldstone boson, the famous QCD axion \cite{Weinberg:1977ma,Wilczek:1977pj}. 
Non-perturbative effects generate a mass for the QCD axion mass which must be less than ${\cal O}(10)$ meV to satisfy the current astrophysical observational 
bounds~\cite{Hamaguchi:2018oqw,Beznogov:2018fda,Leinson:2019cqv}.
Going beyond QCD, axion-like particles arise in string theory \cite{Arvanitaki:2009fg} and can solve other open questions in the Standard Model of particle physics
(SM) such as the hierarchy problem~\cite{Graham:2015cka}, 
be responsible for non-thermal dark matter (DM) via vacuum misalignment mechanism \cite{Preskill:1982cy,
Abbott:1982af, Dine:1982ah}, and may also account for the dark energy of the Universe \cite{Jain:2004gi,
Kim:2009cp, Kim:2013jka, Lloyd-Stubbs:2018ouj} and lead to baryogenesis as a solution to the matter antimatter asymmetry puzzle \cite{Daido:2015gqa, DeSimone:2016bok}. It has also been recently considered in the context of axion-driven kination which is like a stiff equation of state in pre-BBN era
\cite{Co:2019jts,Co:2019wyp,Co:2020jtv,Harigaya:2021txz,Co:2021qgl}. 


The paper is organized as follows: In Section~\ref{sec_genfor}, we discuss PBH formation in curvaton scenario and the curvaton's dynamics during and after inflation. In Section~\ref{sec_axionfields}, we discuss the ALP model, and in Section~\ref{sec:analytical_solutions}, we develop the analytical treatment of the ALP curvaton dynamics. In Section~\ref{sec:benchmark_points}, we discuss the parameter space through benchmark points. Section~\ref{sec:particle_dark_matter_scenario} is reserved for extended scenarios and PBH clustering, and Section~\ref{sec:discussion_and_conclusions} contains a discussion and conclusions. Unless otherwise noted, we use natural units with $c=\hbar=k_B=\MPl=1$.

\medskip

\section{Primordial black holes from curvaton scenarios: the main idea}
\label{sec_genfor}

The curvaton is a scalar field that is subdominant during inflation but contributes to curvature perturbations after inflation has ended.
The curvaton scenario was introduced in \cite{Lyth:2001nq,Moroi:2001ct,Enqvist:2001zp} to
``liberate'' the inflaton field from being responsible for producing curvature fluctuations at large CMB scales (see e.g., \cite{Dimopoulos:2002kt}). Curvaton models enlarge the set of inflationary scenarios whose predictions are compatible with data, and improve fine-tuning issues affecting  several inflationary setups. Moreover, after their introduction, these models led to the development of several interesting ideas based on the dynamics of spectator fields during inflation (see e.g., \cite{Bassett:2005xm} for a review).

A perspective related to the curvaton mechanism has been recently developed in \cite{Stamou:2023vft, Stamou:2023vwz} in the context
of PBH model building (see also \cite{Carr:2019hud, Hardwick:2017fjo,Gow:2023zzp} for related constructions). 
The authors consider inflationary models
where the dynamics of a spectator field during inflation leads to controllable PBH production in the early Universe, with reduced fine-tuning in the parameters. This favourable condition can be obtained by exploiting the strong non-Gaussian tails of curvature perturbations, induced by the spectator field dynamics after inflation ends.

In this Section, we describe the ideas at the basis of the PBH-curvaton scenario  of \cite{Carr:2019hud,Stamou:2023vft, Stamou:2023vwz, Stamou:2024xkk}, focusing on the dynamics of cosmological perturbations.
The  perturbations in the curvaton field, created during inflation, are translated into curvature perturbations after inflation ends. 
Calculations can be carried out in a transparent way by means of the $\delta\mathcal{N}$ approach~\cite{Sasaki:1995aw}. Different parts of the Universe expand by different amounts, as the curvaton field---denoted by $\psi(x)$---rolls through its potential towards a final hypersurface of fixed energy density. If the potential has localised features, the small-scale curvature perturbations arising around the features get amplified, acquiring pronounced non-Gaussian statistics. This leads to PBH formation at small scales. We aim to build a curvaton PBH scenario where the effects of the curvaton $\psi$ are negligible at large cosmic scales, hence CMB and LSS observations are not affected.

Following \cite{Stamou:2023vft, Stamou:2023vwz}, we start by considering our observable Universe at a very early time, when its  size is of the order of the Hubble radius. We measure time during inflation in terms of e-folds of expansion, call them $N$, normalized so that at this initial time $N=0$. We denote the (average) curvaton field value in our Hubble patch at this initial time with $\psi_0$. 

As  time proceeds forward, cosmic expansion stretches the patch that corresponds to our observed Universe: soon it covers many Hubble-sized sub-regions. Each of these regions can be treated as a ``separate Universe'' \cite{Wands:2000dp} with its own, locally homogeneous field value. In each patch, the curvaton's classical motion is frozen, due to the Hubble friction associated to the background expansion. However, the curvaton field experiences stochastic-type diffusion, due to new  modes constantly produced by quantum effects of the Universe's expansion (see e.g. \cite{Vennin:2020kng}). During inflation, the local curvaton $\psi$ then follows a stochastic equation of motion
\begin{equation} \label{eq:curvaton_Langevin_equation}
    \dd \psi = \sigma_N\sqrt{\dd N}\xi_N \, , \qquad
    \sigma_N \equiv \frac{H_*}{2\pi} \, , \qquad
    \expval{\xi_N \xi_{N'}} = \delta(N-N') \, , 
\end{equation}
where $H_*$ is the Hubble parameter during inflation, which we take to be approximately constant\footnote{In \cite{Stamou:2023vft, Stamou:2023vwz, Stamou:2024xkk}, the Hubble parameter has a mild scale dependence during inflation, compatible with the CMB spectral tilt. We expect such a scale dependence to only have a small effect in our considerations, hence for simplicity we omit it. Explicitly, for a scale dependence of the form $\sigma_N = \frac{H_*}{2\pi}e^{-\epsilon_1 (N-N_*)}
$ where $\epsilon_1$ is a small, approximately constant first slow-roll parameter, we get $\Sigma_N^2 = \frac{H_*^2}{4\pi^2}\frac{e^{2\epsilon_1 N_*} - e^{-2\epsilon_1 (N-N_*)}}{2\epsilon_1} = \frac{H_*^2}{4\pi^2}\qty(N + \mathcal{O}(N^2\epsilon_1))$. For $N\sim\mathcal{O}(10)$ and $\epsilon_1 \lesssim \mathcal{O}(10^{-3})$ (the largest value allowed by constraints on the tensor-to-scalar ratio), the correction is negligible.}, while $\xi_N$ is a white, Gaussian noise controlled by the quantum effects mentioned above. Assuming inflation is driven by a single field separate from the curvaton, CMB results on the tensor-to-scalar ratio $8 P_{T}/P_{\zeta} < 0.036$ and scalar power spectrum $A_s \approx 2.1 \times 10^{-9}$ \cite{Planck:2018jri, BICEP:2021xfz} imply
\begin{equation} \label{eq:CMB_constraints}
    H_* < H_\text{max} = 4.7 \times 10^{13} \, \text{GeV} \, .
\end{equation}
 
Solving Eq.~\eqref{eq:curvaton_Langevin_equation}, we
deduce that the curvaton field in a given  Hubble patch at time $N$ obeys a Gaussian probability distribution
\begin{equation} \label{eq:curvaton_distribution}
    P(\psi,N) = \frac{1}{\sqrt{2\pi}\Sigma_N}e^{-\frac{(\psi - \psi_0)^2}{2\Sigma_N^2}} \, , \qquad
    \Sigma_N^2 \equiv \int_0^N \dd N' \sigma_{N'}^2 = \frac{H_*^2}{4\pi^2}N \, .
\end{equation}
The distribution's width $\Sigma_N^2$ is proportional to $N$, even when $\sigma_N$ is constant---and this is the main source of scale-dependence in our model.

Equation \eqref{eq:curvaton_distribution} can be interpreted as the probability distribution for the coarse-grained curvaton $\psi$ containing all Fourier modes up to a scale $k \propto e^N H_*$ \cite{Figueroa:2020jkf}. In position space\footnote{To obtain the white-noise stochastic equations~\eqref{eq:curvaton_Langevin_equation}, the coarse graining is performed in terms  of a step function in Fourier space. In real space, the corresponding window function oscillates with a decaying amplitude. We omit this complication for simplicity and make the identification $r=1/k$. For a more accurate treatment, see e.g.~\cite{Raatikainen:2023bzk}.}, $\psi$ represents the average of the field in a patch of comoving size $r$. These scales are related by 
\begin{equation} \label{eq:N_vs_k_vs_r}
\dd \ln r = -\dd \ln k = -\dd N\,, \qquad r = \frac{1}{k} = \frac{1}{k_*}e^{N_*-N} \, .
\end{equation}
We fix the proportionality at the CMB pivot scale $k_*=\SI{0.05}{Mpc^ {-1}}$. Taking the comoving size of the observable Universe to be the inverse of the Hubble parameter today, and choosing $H_0 = \SI{70} {km/s/Mpc}$, we obtain $N_* \approx \ln k_*/H_0 \approx 5$.
Below, we use the quantities $N$, $k$, and $r$ interchangeably without ambiguities. Each scale also corresponds to a specific PBH mass, 
as we will discuss in Section~\ref{sec:PBH_numbers}.


Hubble friction freezes the curvaton perturbations after they leave the Hubble horizon. We denote the frozen super-horizon value of the curvaton field by $\tilde{\psi}$. Coarse-grained over a distance $r$, its probability distribution $P(\tilde{\psi},r)$ is given by Eq.~\eqref{eq:curvaton_distribution} with the identification \eqref{eq:N_vs_k_vs_r}.
Curvaton perturbations start to evolve again during the post-inflationary radiation-dominated epoch. In \cite{Carr:2019hud,Stamou:2023vft, Stamou:2023vwz, Stamou:2024xkk}, the authors consider a scenario where the curvaton briefly becomes the dominant energy density component. In our work, instead, we focus on scenarios where the curvaton contribution to energy density is {\it always subdominant} with respect to radiation. There maybe several instances where such a sub-dominant curvaton or spectator field field could be interesting: firstly it does not decay this may remain itself as the dominant component the dark matter of the Universe \cite{Tenkanen:2019aij,Markkanen:2018gcw}. Moreover sub-dominant curvaton naturally leads to larger non-Gaussian tails which means significant PBH formation at small scales \cite{Sasaki:2006kq}. Finally there ar several well motivated particle physics based scenarios where the inflationary reheating is left over with sub-dominant energy density in the dark sector (assuming inflationary reheating transfers its energy density into visible sector), (see Ref. \cite{Ghoshal:2022fud} and references there-in).
Nevertheless, even when this condition is satisfied, its impact on the expansion of the Universe becomes non-negligible during radiation domination. 
During this post-inflationary epoch, we denote with $N_p$ the expansion e-folds (setting $N_p=0$ at $\rho_r = \rho_\text{dec}$, see below);  the equations of motion governing the curvaton evolution are
\begin{equation} \label{eq:curvaton_radiation_eoms}
    \psi'' + \qty[\qty(3-\frac{1}{2}\psi'^2) - \frac{2}{3}\frac{\rho_r}{H^2} ]\psi' + \frac{V'}{H^2} = 0 \, , \quad
    H^2 = \frac{V + \rho_r}{3 - \frac{1}{2}\psi'^2} \, , \quad
    \rho_r = \rho_\text{dec}e^{-4N_p} \, ,
\end{equation}
where a prime denotes a derivative with respect to $N_p$, and $\rho_r$ is the radiation energy density\footnote{In \cite{Stamou:2023vft, Stamou:2023vwz, Stamou:2024xkk}, the $\psi'$ dependence in the $H^2$ term is neglected. For completeness, we include its contribution in our analysis. Equation \eqref{eq:curvaton_radiation_eoms} is general and applies, in particular, for both radiation and curvaton domination.}. The evolution of the curvaton field and its energy density depends on the curvaton potential $V(\psi)$. We  solve these equations starting from a given curvaton value  with vanishing initial velocity. We follow the system evolution up to a hypersurface of constant {total} energy density, $\rho=\rho_\text{dec}$, corresponding to the stage of curvaton decay. 
The e-fold number when the process completes is given by a function $\tilde{N}(\tilde{\psi})$, relating each initial, perturbed curvaton value $\tilde{\psi}$ to an e-fold value $\tilde{N}$. 
The $\delta\mathcal{N}$ formalism connects the curvature perturbation in a patch of space to $\tilde{\psi}$ as \cite{Sasaki:1995aw, Sasaki:1998ug, Wands:2000dp, Lyth:2004gb}
\begin{equation} \label{eq:delta_N_formula}
    \zeta = \tilde{N}(\tilde{\psi}) - \bar{N} \, , \qquad \bar{N} \approx \tilde{N}(\psi_0) \, .
\end{equation}
For a coarse-grained $\psi$, we interpret the quantity $\zeta$ above as a coarse-grained curvature perturbation.\footnote{For coarse-grained, super-Hubble quantities, it would arguably be more accurate to make a comparison to the average $\tilde{N}$ that the field takes in all the Hubble-sized sub-patches, see e.g.,~\cite{Tada:2021zzj}. However, this complicates the computation considerably, and we prefer not to follow this route.}

\begin{figure}
    \centering
    \includegraphics{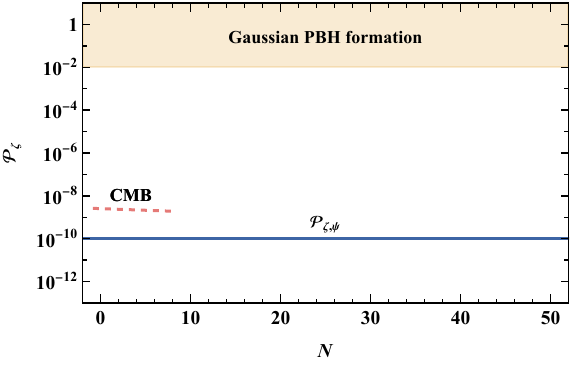}
    \caption{\it The curvaton power spectrum contribution~\eqref{eq:power_spectrum_from_curvaton} for model~A from Table~\ref{tab:benchmarks} (blue), compared to the CMB observations \cite{Planck:2018jri} (dashed red) and the Gaussian PBH formation threshold of $\Pp_\zeta \sim 10^{-2}$ (shaded).}
    \label{fig:power_spectrum}
\end{figure}

In the limit of small, linear perturbations, we can Taylor expand Eq.~\eqref{eq:delta_N_formula} around the value $\tilde{\psi}=\psi_0$,
obtaining\footnote{We use lower indices to indicate derivatives, so that ${N}_{\tilde{\psi}} \equiv \dd N / \dd \tilde{\psi}$, and, in a slight abuse of notation, $\tilde{N}_{\psi_0} \equiv \left(  {\dd N}/{\dd \tilde{\psi}} \right)\big|_{\tilde{\psi}=\psi_0}$.}  $\zeta \approx \tilde{N}_{\psi_0}\delta\tilde{\psi}$. The curvature power spectrum associated
with the curvaton field is then
\begin{equation} \label{eq:power_spectrum_from_curvaton}
    \Pp_{\zeta,\psi}(k) = \Pp_{\psi}(k) \tilde{N}_{\psi_0}^2 = \frac{H_*^2}{4\pi^2} \tilde{N}_{\psi_0}^2 \, ,
\end{equation}
where in the second equality we use the curvaton statistics arising from the process~\eqref{eq:curvaton_Langevin_equation}. Hence, the properties of the curvature power spectrum depend on the first derivative of the e-fold number along the direction of the curvaton in field space. Note that the power spectrum is scale-independent due to the scale-independence of Eq.~\eqref{eq:curvaton_Langevin_equation}. We make it subdominant to the inflaton contribution, so it does
not affect  the CMB constraints. See Fig.~\ref{fig:power_spectrum} for a
pictorial representation of the situation we are considering.

Conventionally, an analysis of PBH formation is developed in terms of conditions on the power spectrum $\Pp_\zeta$, demanding that it becomes larger than a threshold of order $10^{-2}$ at small scales, where the scale is related to the PBH mass, see e.g.,~\cite{Karam:2022nym}. Since PBHs form from large fluctuations in the distribution's tails, non-Gaussianities are often important; a more careful analyses sometimes work with the local value of the curvature perturbation $\zeta$ computed beyond Gaussian order, see e.g.,~\cite{Tomberg:2023kli}. However, it has been argued~\cite{Young:2014ana} that one should consider only the local value of $\zeta$, subtracting the effect of long-wavelength modes, since these do not affect the gravitational collapse. In \cite{Stamou:2023vft}, the authors took this route and considered an ``inner'' perturbation in $\zeta$ on top of the local background. We follow the same route, but we connect the ``inner'' perturbation to the perturbation  \emph{compaction function}.

Numerical studies show that PBH formation is best estimated in terms of the perturbation  compaction function \cite{Shibata:1999zs,Musco:2018rwt, Escriva:2021aeh}, the mass excess inside a region of space divided by the radius:
\begin{equation} \label{eq:compaction_defined}
    \Cc \equiv \frac{2 \delta M}{R} \, .
\end{equation}
If this quantity is of order one, the mass is  sufficiently concentrated (essentially, inside the corresponding Schwarzschild radius) for a black hole to form.
In the super-Hubble limit, assuming spherical symmetry, the compaction function can be written in terms of the radial derivative of the curvature perturbation,
\begin{equation} \label{eq:compaction_function_in_zeta}
    \Cc = \frac{2}{3}\qty[1-\qty(1+\frac{\dd \zeta(r)}{\dd \ln r})^2] \, .
\end{equation}
The collapse threshold is $\Cc_\text{th} \approx 0.4$ (we neglect, for simplicity, the shape-dependence of the threshold). For convenience, we define the ``linear'' (and rescaled) compaction $\Cc_l \equiv {\dd \zeta(r)}/{\dd \ln r}$; the collapse criterion can equivalently be written as $\Cc_l < \Cc_{l,\text{th}} \approx -0.37$. Let us emphasize that this criterion applies for super-Hubble perturbations; below, we compute $\Cc_l$ right after the curvaton decay, after which the superhorizon curvature perturbations (and thus $\Cc$) stay frozen, and we consider PBHs formed at scales that re-enter the Hubble radius after the decay.

To estimate the radial derivative in Eq.~\eqref{eq:compaction_function_in_zeta}, 
we consider e-fold variations of order one, the typical time-scale of evolution during inflation, and thus set $|\dd N| = \dd \ln r = 1$. This corresponds to assuming that the perturbations are correlated over scales roughly one e-fold apart.\footnote{A similar assumption is present in most simple estimates of the PBH abundance: in \cite{Stamou:2023vft, Stamou:2023vwz} on which our formalism is based, and also when PBHs from logarithmic mass bins are assumed to form independently, see, e.g., \cite{Yoo:2022mzl, Escriva:2022duf} and the discussion around equation \eqref{eq:f_PBH_tot}. To drop such an assumption requires a complicated numerical analysis which is beyond the scope of this paper---see, e.g., \cite{Raatikainen:2023bzk}. We examine the assumption in more detail in Section~\ref{sec:comments_on_approximations}.} Using the $\delta\mathcal{N}$ approach, we compute $\dd \zeta \approx \tilde{N}_{\psi} \dd \tilde{\psi}$, see Eq.~\eqref{eq:delta_N_formula}, where $\dd \tilde{\psi}$ has a Gaussian distribution with zero mean and variance $\sigma_N^2$, as per Eq.~\eqref{eq:curvaton_Langevin_equation}. The derivative $\tilde{N}_{\tilde{\psi}}$ depends on the local value of $\tilde{\psi}$. Hence, to build a probability distribution for $\Cc_l$, we integrate over all possible values of the local curvaton field \cite{Stamou:2023vft}:
\begin{equation} \label{eq:Ccl_distribution}
    P(\Cc_l,r) = \int \dd \tilde{\psi} P(\tilde{\psi},r) \frac{1}{\sqrt{2\pi}\sigma_r|\tilde{N}_{\tilde{\psi}}|}e^{-\frac{\Cc_l^2} {2\sigma^2_r\tilde{N}_{\tilde{\psi}}^2}} \, .
\end{equation}
To recap, in \eqref{eq:Ccl_distribution}, $\tilde{N}_{\tilde{\psi}}$ depends on $\tilde{\psi}$, which we marginalize out by integrating over its probability distribution \eqref{eq:curvaton_distribution}. We have exchanged the $N$-dependence of the distribution to an $r$-dependence by the equivalence \eqref{eq:N_vs_k_vs_r}. We have also switched $N$ for $r$ in $\sigma_r \equiv \sigma_N$ from \eqref{eq:curvaton_Langevin_equation} (in our case, this is a constant with no scale dependence).

The result is, in general, non-Gaussian in $\Cc_l$. Such non-Gaussianity  plays an important role in our analysis. To obtain the PBH abundance, denoted with $\beta$,  we integrate the distribution over the collapse region $\Cc_l < \Cc_{l,\text{th}}$:
\begin{equation}
    \beta_N = \int_{-\infty}^{\Cc_{l,\text{th}}} \dd \Cc_l \, P(\Cc_l,N) \, .
\end{equation}
This is sensitive to the non-Gaussian tails of the probability function $ P(\Cc_l,N)$. 



These are the general formulas we use below to study example models. The scale-dependence of the perturbations is important: the models must both produce PBHs at small scales and at the same time not violate the CMB constraints at large scales.
To see how this is possible with the distribution \eqref{eq:Ccl_distribution}, note that if the variance $\tilde{N}_{\tilde{\psi}}^2$ is large, high values of $|\Cc_l|$ become likely, triggering PBH production. As we will see below, $\tilde{N}_{\tilde{\psi}}$ grows near a local hilltop in the curvaton potential. With the correct initial conditions, the probability distribution $P(\tilde{\psi},r)$ supports such field values only for small $r$, which have had more time to diffuse to the hilltop, corresponding to a larger $\Sigma_N$ in \eqref{eq:curvaton_distribution}, leaving large scales and the CMB untouched. We also need to avoid generating too many small-mass PBHs, which would lead to an excessive amount of Hawking radiation and violate the Big Bang Nucleosynthesis bounds, see e.g.,~\cite{Carr:2020gox}. Below, we will see how to address these issues,  and restrict the PBHs into a specific mass window, such as the asteroid mass scale.


\section{Axion-like fields}
\label{sec_axionfields}




As our example model, we consider cosmological aspects of 
ALPs (see e.g., \cite{Marsh:2015xka} for a review) as curvaton candidate. An ALP arises from a complex scalar field $\Phi$, whose potential only depends on the modulus $|\Phi|$. At some high energy scale denoted by $f_a$, the potential develops a minimum at $|\Phi| = f_a$, and the $U(1)$ symmetry related to the phase shifts of $\Phi$ (called the Peccei--Quinn symmetry) is spontaneously broken.\footnote{This symmetry breaking creates axion strings. In our setup, they are diluted by inflation.}
We decompose the field as
\begin{align} \label{eq:phix}
\Phi(x) \ = \ \frac{1}{\sqrt 2}\left[f_a+\sigma(x)\right]e^{i\psi(x)/f_a} \, .
\end{align}
The modulus $\sigma$ has a large mass $m_\sigma\sim f_a$ and can be integrated out, while the angular part $\psi$ becomes the massless pseudo-Nambu Goldstone boson. We assume $f_a > H_*$, so that the Peccei-Quinn symmetry is broken before inflation starts; $\psi$ evolves stochastically and becomes the curvaton of the previous Section. This is the ``pre-inflationary axion'' scenario.\footnote{Contrary to this study, most of the early works considering axion or ALP as the curvaton either used $\sigma$ as the curvaton, or used $\psi$ but only considered the small-field regime \cite{Kawasaki:2011pd,Kawasaki:2012wr,Kawasaki:2012gg,
Ando:2017veq,Ando:2018nge,Kawasaki:2021ycf,Sasaki:1995aw}.}

\begin{figure}
    \centering
    \includegraphics{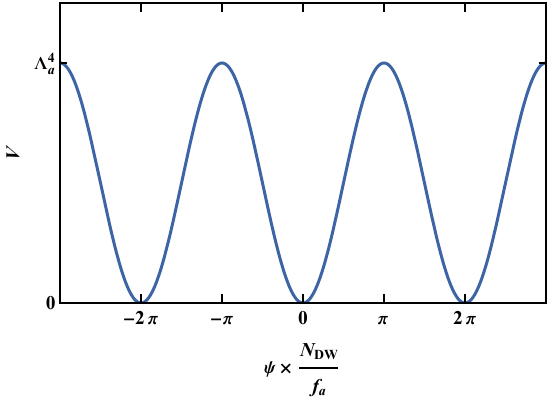}
    \caption{\it A sketch of the axion-like potential \eqref{eq:V_axion}.}
    \label{fig:axion_potential}
\end{figure}

After inflation, at a lower energy $\Lambda_a$, the $U(1)$ symmetry gets explicitly broken by non-perturbative effects, and the ALP $\psi$ obtains a periodic potential, which we take to be of the standard form \cite{Aldazabal:2015yna, DiLuzio:2020wdo}\footnote{In case of the QCD axion, it is this explicit symmetry breaking which solves the strong CP problem.} 
\begin{equation} \label{eq:V_axion}
    V(\psi) = \Lambda_a^4\qty[1 - \cos(\frac{N_\text{DW}\psi}{f_a})] \, ,
\end{equation}
see Fig.~\ref{fig:axion_potential}.
The domain wall number $N_\text{DW}$
denotes the number of distinct vacua as $\psi$ changes from $0$ to $2\pi f_a$. If $N_\text{DW} > 1$, domain walls may form. If the walls extend over long distances, they may spoil early Universe cosmology by becoming dominant~\cite{DiLuzio:2020wdo}. If they only form around rare, closed patches of space, they form bubbles that may form primordial black holes on their own right~\cite{Ferrer:2018uiu}. We take $N_\text{DW}=1$ below, avoiding such issues.

The ALP mass at the potential minimum is $m_a^2 = V''(0) = \Lambda_a^4/f_a^2$. To be more precise, as we approach the energy $\Lambda_a$, the ALP mass develops in a manner dependent on the temperature $T$ of the Universe,
\begin{equation} \label{eq:mofT}
 m_a (T) \simeq 
 \begin{dcases}
    \frac{\Lambda_a^2}{f_a} \left(\frac{\Lambda}{T}\right)^Q
       & \mathrm{for}\, \, \, T > \Lambda_a \, , \\
    \frac{\Lambda_a^2}{f_a} 
       & \mathrm{for}\, \, \, T < \Lambda_a \, ,
 \end{dcases}
\end{equation}
with a positive power $Q$ of order unity given by non-perturbative estimations in the strongly coupled regime \cite{DiLuzio:2020wdo}. For simplicity, we always use the $T < \Lambda_a$ limit. The ALP potential is then fully described by any two of the three quantities $\Lambda_a$, $f_a$, and $m_a$.

Potential \eqref{eq:V_axion} has all the desired properties mentioned at the end of Section~\ref{sec_genfor}. We take $\psi$ to start its stochastic inflationary evolution on the potential slope at $0 < \psi_0 < \pi f_a$ (this is called the misalignment mechanism); rare fluctuations at large $N$, or small length scales, push $\psi$ to the hilltop at $\psi = \pi f_a$, activating PBH production. As we will explain in Section~\ref{sec:PBH_numbers}, small length scales correspond to small PBH masses. Note, however, that the PBH production can only activate after the ALP potential has developed. This provides a lower cutoff in PBH masses, circumventing the problem of excessive Hawking radiation.

When the axion-curvaton starts to evolve in the post-inflationary Universe, it oscillates around the minimum of its potential, until it decays to SM degrees of freedom. ALPs typically interact with the SM through a Lagrangian term of the form
\begin{equation} \label{eq:L}
\mathcal{L_{\rm int}} = \frac{ \alpha }{8 \pi } \frac{\psi}{f_a} F_{\mu \nu} \tilde{F}^{\mu \nu }
\,,
\end{equation}
where $\alpha$ is the dimensioless  coupling between the ALP $\psi$ and a CP-violating Chern-Simons--like term of the electromagnetic field strength $F$.
At temperatures below $\Lambda_a$, the associated decay width is
\begin{equation} \label{eq:Gamma_a}
 \Gamma_a \approx \frac{\alpha^2}{256 \pi^3} \frac{\Lambda_a^6}{f_a ^5} \, .
\end{equation}
We assume instantaneous decay when $H = \Gamma_a$, at energy density $\rho_\text{dec} = 3\MPl \Gamma_a^2$ introduced in Eq.~\eqref{eq:curvaton_radiation_eoms}. Adjusting $\alpha$ in Eq.~\eqref{eq:Gamma_a} lets us set $\rho_\text{dec}$ freely.\footnote{In the benchmark models of Table~\ref{tab:benchmarks}, $\alpha$ is large, many orders of magnitude above one. This is a consequence of the large separation between $\Lambda_a$ and $f_a$. A perturbative treatment of the decay is questionable in this case. We choose to remain agnostic of the details of the decay process and treat $\rho_\text{dec}$ as a free parameter in our model. See also \cite{DiLuzio:2020wdo}.} In Section~\ref{sec:particle_dark_matter_scenario}, we briefly consider a scenario where the ALP does not decay.

In the next Section, we analyse the ALP's post-inflationary dynamics in detail to compute the model's PBH statistics.

\section{Analytical solution for axion-like particles}
\label{sec:analytical_solutions}


We discussed  in Section \ref{sec_genfor}  the general ideas and formulas at the basis of our PBH curvaton scenario, while in Section \ref{sec_axionfields}, we outlined a possible realization of these ideas in terms of ALPs. In this Section, we explicitly solve the curvaton evolution equations in the case of ALPs, and we analyze their ramifications for the statistics of the compaction function. Consequences for PBH production are discussed in Section \ref{sec:benchmark_points}. 

%
%
%
In order to handle more easily the evolution of the ALP,
 we define the rescaled variables
\begin{equation} \label{eq:scaled_variables}
    \theta \equiv \frac{\psi}{f_a} \, , \quad
    g(\theta) \equiv 1-\cos(\theta) = \frac{V(\psi)}{\Lambda_a^4} \, , \quad
    R \equiv \frac{\rho_r}{\Lambda_a^4} \, , \quad
    q \equiv \frac{H_*}{2\pi f_a} \, .
\end{equation}
Similarly, we write $\ttheta \equiv \tilde{\psi}/f_a$ for the frozen super-Hubble $\theta$. In terms of these rescalings, the stochastic equations \eqref{eq:curvaton_Langevin_equation}, \eqref{eq:curvaton_distribution} become
\begin{equation} \label{eq:theta_stochastics}
    \dd \theta = q\sqrt{\dd N}\xi_N \, , \quad P(\theta, N) = \frac{1}{\sqrt{2\pi N}q}e^{-\frac{(\theta-\theta_0)^2}{2q^2N}} \, .
\end{equation}
The post-inflationary equations \eqref{eq:curvaton_radiation_eoms} become
\begin{equation} \label{eq:theta_eom}
    \theta'' + \qty(3-\frac{1}{2}\theta'^2f_a^2)\qty(\qty[1-\frac{2}{3}\frac{R}{R+g(\theta)}]\theta' + \frac{g'(\theta)}{\qty[R+g(\theta)]f_a^2}) = 0 \, , \quad
    R = R_\text{dec} e^{-4N_p} \, .
\end{equation}
We follow the normalization of the post-inflationary
e-fold number $N_p$ introduced in Eq.~\eqref{eq:curvaton_radiation_eoms}, defining the rescaled quantity $R_\text{dec} \equiv \rho_\text{dec}/\Lambda_a^4$.

The curvature power spectrum \eqref{eq:power_spectrum_from_curvaton} reads
\begin{equation} \label{eq:power_spectrum_theta}
    \Pp_{\zeta,\psi}(k) = q^2 \tilde{N}_{\theta_0}^2 \, , \qquad \theta_0 \equiv \frac{\psi_0}{f_a} \, ,
\end{equation}
where $\tilde{N}_{\theta}$ is the $\theta$-derivative of $\tilde{N}$ and $\tilde{N}_{\theta_0} \equiv \tilde{N}_{\theta}(\theta_0)$. The non-Gaussian probability distribution for the linear compaction function $\Cc_l$ \eqref{eq:Ccl_distribution} results in
\begin{equation} \label{eq:Ccl_distribution_scaled}
    P(\Cc_l,N) = \frac{1}{2\pi q^2 \sqrt{N}}\int \frac{\dd \ttheta}{|\tilde{N}_{\ttheta}|} \exp(-\frac{(\ttheta-\theta_0)^2}{2q^2N} - \frac{\Cc_l^2}{2q^2\tilde{N}_{\ttheta}^2}) \, .
\end{equation}
We now proceed to solve Eq.~\eqref{eq:theta_eom} analytically in different regimes.

\subsection{Near the hilltop}
\label{sec:analytics_near_hilltop}

\begin{figure}
    \centering
    \includegraphics{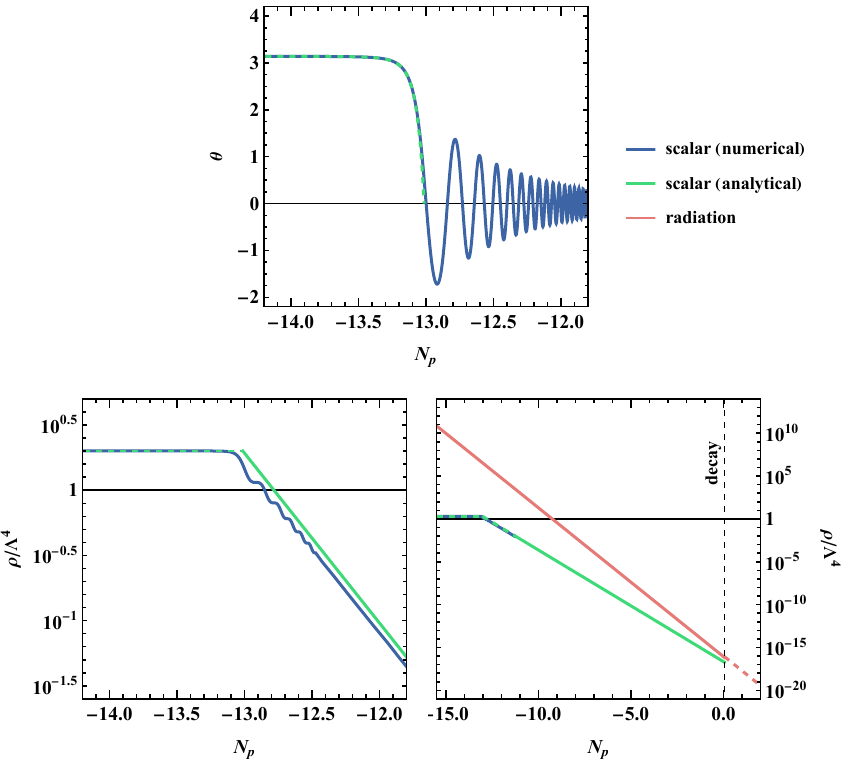}
    \caption{\it Evolution of $\theta$ and its energy density, together with the energy density of radiation, from the initial value $\ttheta_\text{pk}$, in model~A from Table~\ref{tab:benchmarks}. The numerical curves correspond
    to solutions of Eq.~\eqref{eq:theta_eom}. The analytical solution for $\theta$ follows Eq.~\eqref{eq:phi_solution} up to $N_\text{osc}$
     given by Eq.~\eqref{eq:n_osc_in_phi_ini}. The analytical solution for the scalar energy density is frozen at $g(\ttheta_\text{pk})$ until $N_\text{osc}$, and then starts to scale as non-relativistic  matter. Such a behaviour  slightly overestimates the late-time energy density.}
    \label{fig:theta_evolution}
\end{figure}

We solve Eq.~\eqref{eq:theta_eom} in the limit $\theta \to \pi$  near the hilltop of
the curvaton potential, focusing on the case of radiation domination, $R \gg g(\theta)$. In doing this, our
aim is  to explore the high-$\Cc_l$ tail of the probability distribution~\eqref{eq:Ccl_distribution_scaled}. Defining the new variable $\varphi \equiv \pi - \theta$, we get $g(\theta) \approx 2 - \frac{1}{2}\varphi^2$, and so
\begin{equation} \label{eq:phi_approx_eom}
    \varphi'' + \varphi' - \frac{3\varphi}{f_a^2R_\text{dec}}e^{4N_p} \approx 0 \, .
\end{equation}
We start around the hilltop with zero field velocity, dropping the non-linear $\theta'$ contribution in Eq.~\eqref{eq:theta_eom}.
We can further scale away the effects of the $f_a$ and $R_\text{dec}$ parameters by defining the new time variable 
\begin{equation} \label{eq:n_def}
n \equiv -\frac{1}{4}\ln[R(N_p) f_a^2] = N_p - \frac{1}{4}\ln R_\text{dec} - \frac{1}{2}\ln f_a
\,.
\end{equation}
 Hence, we obtain
\begin{equation} \label{eq:phi_approx_eom_2}
    \varphi'' + \varphi' - 3\varphi e^{4n} \approx 0 \, .
\end{equation}
This equation is solved  in terms of the Bessel functions, imposing the initial conditions $\varphi \to \tilde{\varphi}$, $\varphi' \to 0$ at $n \to -\infty$:
\begin{equation} \label{eq:phi_solution}
\begin{aligned}
    \varphi(n) &= \tilde{\varphi} \times \frac{\sqrt{2}}{3^{1/8}}\Gamma\qty(\frac{5}{4}) e^{-\frac{n}{2}} I_{1/4}\qty(\frac{\sqrt{3}}{2}e^{2n}) \\
    &\approx \tilde{\varphi} \times \sqrt{\frac{2}{\pi}}\frac{1}{3^{3/8}} \Gamma\qty(\frac{5}{4})\exp(\frac{\sqrt{3}}{2}e^{2n}-\frac{3n}{2})  \, .
\end{aligned}
\end{equation}
The last approximation applies specifically to the large-$n$ limit.
In this large-$n$ regime, though, our approximations start to break down: the curvaton leaves the region of the hilltop and starts to oscillate around the minimum of its potential. 
We can approximate this transition to occur sharply, at the point the expression \eqref{eq:phi_solution} satisfies $\varphi(n_{\rm osc}) = \pi$ (in other words, $\theta$ crosses zero).
\footnote{Note that the onset of curvaton oscillation for our non-quadratic potential differs from the definition used in some papers, such as Ref.~\cite{Kawasaki:2011pd}. However, Fig.~\ref{fig:theta_evolution} demonstrates that our analytical approximation closely matches the numerical data.} 
Using the second line of Eq.~\eqref{eq:phi_solution}, this condition gives for the initial oscillation time
\begin{gather}
    \label{eq:phi_ini_in_n_osc}
    \tilde{\varphi} \approx \frac{\pi^{3/2} 3^{3/8}}{\sqrt{2}\Gamma\qty(\frac{5}{4})}\exp(-\frac{\sqrt{3}}{2}e^{2n_\text{osc}}+\frac{3n_\text{osc}}{2}) \Leftrightarrow
    \\
    \label{eq:n_osc_in_phi_ini}
    n_\text{osc} \approx
    -\frac{2}{3}\ln(\frac{\pi^{3/2} 3^{3/8}}{\sqrt{2}\tilde{\varphi}\Gamma\qty(\frac{5}{4})})
    -\frac{1}{2}W_{-1}\qty(-\frac{2^{5/3}\qty[\tilde{\varphi}\Gamma\qty(\frac{5}{4})]^{4/3}}{3\pi^2}) \, ,
\end{gather}
where $W_{-1}$ is the $(-1)$ branch of the Lambert $W$ function.
\footnote{Since we focus on the hilltop, the argument of the Lambert $W$ function in Eq.~\eqref{eq:n_osc_in_phi_ini} is between $-1/e$ and $0$, so the $W_{-1}$ branch gives a real and positive result. See, e.g., \cite{olver2010nist} for the detailed properties of the Lambert $W$ function.}
Note that $n_\text{osc} $ is a function of $\tilde{\varphi}$ \emph{only}: we have scaled away all other parameter dependence.

The curvaton decays when the total scaled energy density of the Universe equals $R_\text{dec}$. We aim to solve for $\tilde{N}$, i.e., the number of e-folds at decay as defined in Eq.~\eqref{eq:delta_N_formula}. After the curvaton starts to oscillate, its energy density scales as cold matter, $\propto e^{-3N_p}$, so we solve  equation
\begin{equation} \label{eq:decay_N_eq}
    R(\tilde{N}) + g(\ttheta)e^{-3(\tilde{N}-N_{p,\text{osc}})} \approx R_\text{dec} \, .
\end{equation}
We take
 $\ttheta = \pi - \tilde{\varphi}$, and note that $g(\ttheta) \approx 2$ near the hilltop. Converting $N_p$ to the time variable $n$, Eq.~\eqref{eq:decay_N_eq} becomes
\begin{equation} \label{eq:decay_N_eq_2}
    e^{-4\tilde{N}} + \frac{2f_a^{3/2}}{R^{1/4}_\text{dec}}e^{-3(\tilde{N} - n_\text{osc})} = 1 \, .
\end{equation}

This is a fourth-order polynomial equation in $e^{-\tilde{N}}$; it can be solved analytically, but the result is quite complex. In our scenario, the curvaton is subdominant at the time of the decay, so we can expand to leading order in both $\tilde{N}$ and $f_a^{3/2}/R_\text{dec}^{1/4}$ to obtain
\begin{equation} \label{eq:Ntilde_approx_3}
    \tilde{N} \approx \frac{f_a^{3/2}}{2R^{1/4}_\text{dec}}e^{3n_\text{osc}} \, .
\end{equation}
Together with Eq.~\eqref{eq:n_osc_in_phi_ini}, this condition provides the function $\tilde{N}(\ttheta)$.

What is its asymptotic behaviour of Eq.~\eqref{eq:Ntilde_approx_3} for small $\tilde{\varphi}$, or equivalently, large $n_\text{osc}$ and $\tilde{N}$? According to Eq.~\eqref{eq:phi_ini_in_n_osc}, $\tilde{\varphi}$ behaves roughly like the double logarithm of $n_\text{osc}$---that is, it changes slowly. At least locally,  we may expect
\begin{equation} \label{eq:Ntilde_scaling}
    \tilde{N} = a - b\ln \tilde{\varphi} \, ,
\end{equation}
where $a$ and $b$ are constants and $b>0$. Hence, $\tilde{N}_{\tilde{\varphi}} = -b/\tilde{\varphi}$. 
Below, we are going to need the value of $b$ around the point that satisfies $\tilde{N}_{\ttheta} = -\tilde{N}_{\tilde{\varphi}} = |\Cc_{l,\text{th}}|/q$ (for reasons explained below, we call the corresponding field value $\ttheta_\text{pk}$). It can be obtained by keeping only the leading exponential term in the exponent of Eq.~\eqref{eq:phi_ini_in_n_osc} and differentiating  both sides with respect to $\tilde{\varphi}$. We also differentiate both sides of Eq.~\eqref{eq:Ntilde_approx_3}.
We solve this pair of equations with the added condition of $\tilde{N}_{\ttheta} = |\Cc_{l,\text{th}}|/q$ and obtain
\begin{equation} \label{eq:b}
    b = \tilde{\varphi} \tilde{N}_{\ttheta} = \frac{3^{1/4}f_a^{3/2}}{2\sqrt{2} R_\text{dec}^{1/4}} \times \sqrt{-W_{-1}\qty(-\frac{3F^6_aq^4\Gamma\qty(\frac{5}{4})^4}{2\Cc^4_{l,\text{th}}\pi^6R_\text{dec}})}
    \qquad
    \text{for} \quad \tilde{N}_{\ttheta} = \frac{|\Cc_{l,\text{th}}|}{q} \, .
\end{equation}
The parameter dependence of $b$ is dominated by the prefactor $\propto f_a^{3/2}R_\text{dec}^{-1/4}$; the Lambert $W$ function is typically of order one, and its dependence on the argument is mild (approximately logarithmic).

Figure~\ref{fig:theta_evolution} shows the behaviour  of the scalar field and the radiation bath, and compares the analytical approximations of this Section to a numerical solution of Eq.~\eqref{eq:theta_eom} in an example case. The solutions for $\theta(N_p)$  agree very well up to the point $\theta = 0$. The energy density approximation, obtained by switching from a constant density to a matter-like scaling at the epoch $n_\text{osc}$, is able to reproduce the numerical solution with  adequate precision.

\begin{figure}
    \centering
    \includegraphics{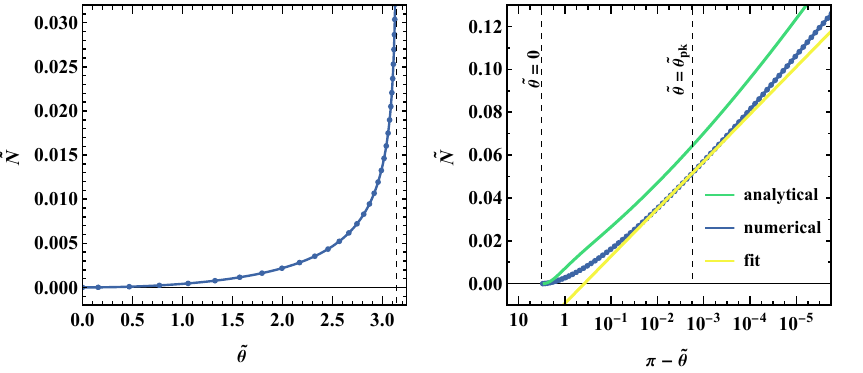}
    \caption{\it The function $\tilde{N}(\ttheta)$ for model~A from Table~\ref{tab:benchmarks}. The numerical line arises from numerically solving Eq.~\eqref{eq:theta_eom}. The analytical curve is given by Eq.~\eqref{eq:Ntilde_approx_3}. It is shifted in the $y$-direction, due to the inaccuracy of approximating $n_\text{osc}$ by Eq.~\eqref{eq:phi_ini_in_n_osc}. The fit is given by Eq.~\eqref{eq:Ntilde_scaling}, with $b$ from Eq.~\eqref{eq:b} and $a$ chosen to coincide with the numerical result at $\ttheta=\ttheta_\text{pk}$.}
    \label{fig:ntilde}
\end{figure}

Figure~\ref{fig:ntilde} represents the functional form of $\tilde{N}(\ttheta)$ in a representative example, comparing  it to Eqs.~\eqref{eq:Ntilde_approx_3} and \eqref{eq:Ntilde_scaling}. Due to the slight inaccuracy of the $n_\text{osc}$ approximation (also apparent in the lower left panel in Fig.~\ref{fig:theta_evolution}) there is a shift between the numerical result and Eq.~\eqref{eq:Ntilde_approx_3} in the $\ttheta \to \pi$ tail; however, this is not relevant for our results for $\Cc_l$ below, which depend mainly on the derivative $\tilde{N}_{\ttheta}$. As the right panel of Fig.~\ref{fig:ntilde} shows, this derivative is captured well by the $b$ factor in Eq.~\eqref{eq:b}.

\subsection{Away from the hilltop}
\label{sec:analytics_away_from_hilltop}

We can also analytically solve the equations well away from the hilltop, in the regime of small $\theta$, and do it here for completeness.  In the small $\theta$ limit, the potential is quadratic, $g(\theta) \approx \frac{1}{2}\theta^2$. The field
variable $\theta$ follows an evolution equation analogous to Eq.~\eqref{eq:phi_approx_eom_2}:
\begin{equation} \label{eq:small_theta_approx_eom}
    \theta'' + \theta' + 3\theta e^{4n} \approx 0 \, ,
\end{equation}
whose solution is 
\begin{equation} \label{eq:small_theta_solution}
    \theta(n) = \ttheta \times \frac{\sqrt{2}}{3^{1/8}}\Gamma\qty(\frac{5}{4}) e^{-\frac{n}{2}} J_{1/4}\qty(\frac{\sqrt{3}}{2}e^{2n}) \, .
\end{equation}
At early times, the solution is frozen at $\theta=\ttheta$. At late times, the solution oscillates around the minimum of the potential, with
\begin{equation} \label{eq:small_theta_oscillations}
    \theta(n) \approx \ttheta \times \frac{2^{3/2}}{3^{3/8}\sqrt{\pi}}\Gamma\qty(\frac{5}{4}) e^{-\frac{3n}{2}} \cos(\frac{\sqrt{3}}{2}e^{2n} - \frac{3\pi}{8}) \, .
\end{equation}
During the oscillations, the energy density scales as $\propto \theta^2 \propto e^{-3n}$, mimicking cold matter as already noted above. From the envelope of the oscillations, setting the cosine to one  and solving the condition $\theta(n_\text{osc}) = \ttheta$ from the ensuing equation, we obtain
\begin{equation} \label{sec:n_osc_small_theta_0}
    n_\text{osc} = -\frac{2}{3}\ln(\frac{3^{3/8}\sqrt{\pi}}{2^{3/2}\Gamma(5/4)}) \approx -0.029
\end{equation}
as the effective ``start time of oscillations''.  $|n_\text{osc}|$ is a small constant independent from $\ttheta$. 
Comparing to Eq.~\eqref{eq:n_def}, we see the oscillations start when $R f_a^2 \approx 1$, giving $3H^2\approx \rho_r = R\Lambda_a^4 = m_a^2$---that is, when $H\sim m_a$, a well-known result widely adopted for a quadratic curvaton, see, e.g., \cite{Lyth:2001nq,Moroi:2001ct,Enqvist:2001zp}.

To obtain $\tilde{N}$, we start again from Eq.~\eqref{eq:decay_N_eq}. This time, $g(\ttheta)=\frac{1}{2}\ttheta^2$ instead of the hilltop value of $2$. We thus get
\begin{equation} \label{eq:decay_N_eq_small_theta}
    e^{-4\tilde{N}} + \frac{\ttheta^2 f_a^{3/2}}{2 R^{1/4}_\text{dec}}e^{-3(\tilde{N} - n_\text{osc})} = 1\,, \quad {\text{hence}}\, \quad
    \tilde{N} \approx \frac{\ttheta^2f_a^{3/2}}{8R^{1/4}_\text{dec}}e^{3n_\text{osc}} \, .
\end{equation}
We can use this result to gain insight into the power spectrum  $\Pp_{\zeta,\psi}$ when $\theta_0$ is small. Naively, the result~\eqref{eq:power_spectrum_theta} implies that $\Pp_{\zeta,\psi} \to 0$ as $\theta_0 \to 0$, since then $\tilde{N}_{\ttheta} \to 0$ (see Fig.~\ref{fig:ntilde})\footnote{Fig.~\ref{fig:ntilde} shows a different shape for the first derivative $\tilde{N}_{\ttheta}$ compared to Fig. 6 in~\cite{Kawasaki:2011pd}, Fig. 1 in~\cite{Kobayashi:2020xhm}, and Fig. 6 in~\cite{Chen:2023lou}. This is due to different conventions: when varying $\ttheta$, we keep constant the energy density at curvaton decay, rather than the energy ratio, as done in previous studies.}, providing an easy way to fulfill the CMB constraints. However, in this limit, the approximation \eqref{eq:power_spectrum_theta} breaks down: $\tilde{N}_{\ttheta}$ is no longer a constant for the ``typical'' perturbations, which have a finite width $q\sqrt{N}$. To estimate the power spectrum in this limit, we can instead use the more general definition (see e.g.,~\cite{Vennin:2015hra})
\begin{equation} \label{eq:power_spectrum_theta_general}
    \Pp_{\zeta,\psi}(k) = \frac{\dd \expval{\tilde{N}^2(N)}}{\dd N} \, ,
\end{equation}
based on the fact that $\expval{\tilde{N}^2(N)}$ equals the integral over the power spectrum up to the scale $N$, as per the $\delta\mathcal{N}$ formalism. Indeed, the spectrum in Eq.~\eqref{eq:power_spectrum_theta} can be derived from Eq.~\eqref{eq:power_spectrum_theta_general} by approximating $\tilde{N} \approx \tilde{N}_{\ttheta} \dd \ttheta$ and using $\expval{\dd \ttheta^2} = q^2 N$.

Using Eq.~\eqref{eq:decay_N_eq_small_theta}, we obtain
\begin{equation} \label{eq:Ntilde_squared_small_theta}
    \expval{\tilde{N}^2} = \frac{f_a^3}{64R^{1/2}_\text{dec}}e^{6n_\text{osc}}\expval{\ttheta^4}
    = \frac{3f_a^3}{64R^{1/2}_\text{dec}}e^{6n_\text{osc}}q^4N^2 \, ,
\end{equation}
where we made use of the Gaussianity of $\ttheta$. We finally get 
\begin{equation} \label{eq:power_spectrum_theta_small}
    \Pp_{\zeta,\psi}(k) = q^4\frac{3\,N f_a^3}{32\,R^{1/2}_\text{dec}}e^{6n_\text{osc}}  \, .
\end{equation}
This is essentially Eq.~\eqref{eq:power_spectrum_theta} evaluated at the ``typical'' values of $\ttheta \sim q\sqrt{N}$. Even for small $\ttheta$, the power spectrum can not be smaller than this value.

After deriving these analytic results on the dynamics
of the curvaton field in different regimes, we
proceed to study their consequences for PBH
production. 

\subsection{Compaction function distribution}
\label{sec:C_distribution}

We are now ready to compute the probability distribution of $\Cc_l$ from Eq.~\eqref{eq:Ccl_distribution_scaled}. At small $|\Cc_l|$, the result may be Gaussian, or instead a more sharply peaked distribution, depending on the value of $\theta_0$, as we show in Appendix~\ref{sec:small_compaction}. In
order to study  PBH production, we are interested in the tail of the distribution at large $|\Cc_l|$.

When $|\Cc_l|$ is large, the first term in the exponent in Eq.~\eqref{eq:Ccl_distribution_scaled} is subleading. Eq.~\eqref{eq:Ntilde_scaling} tells us that
the main contribution to the integral arises from values of $\ttheta$ close to $\pi$, where $\tilde{N}_{\ttheta}$ grows sharply.  We approximate the integral using the saddle point method, writing
\begin{equation} \label{eq:saddle_point_approx}
\begin{gathered}
    \frac{1}{\tilde{N}_{\ttheta}} \exp(-\frac{\Cc_l^2}{2q^2\tilde{N}_{\ttheta}^2}) \equiv \exp[-f(\ttheta)] \approx \exp[-f(\ttheta_\text{pk}) - {1\over2}f''(\ttheta_\text{pk})(\ttheta - \ttheta_\text{pk})^2] \, , \\
    f(\ttheta) = \frac{\Cc_l^2}{2q^2\tilde{N}_{\ttheta}^2} + \ln \tilde{N}_{\ttheta} \, , \quad
    f'(\ttheta_\text{pk}) = 0 \quad \Leftrightarrow \quad \tilde{N}_{\ttheta}(\ttheta_\text{pk}) = \frac{|\Cc_l|}{q} \, , \\
    f''(\ttheta_\text{pk}) = 2\frac{q^2}{\Cc_l^2}\tilde{N}_{\ttheta\ttheta}(\ttheta_\text{pk})^2 \, .
\end{gathered}
\end{equation}
The quantity $\ttheta_\text{pk}$ can be obtained once the function $\tilde{N}$ is known. The Gaussian integral over $\ttheta$ can then be performed, to yield
\begin{equation} \label{eq:Ccl_distribution_tail}
    P(\Cc_l,N) \approx \frac{1}{2 q^2\sqrt{N\pi}\tilde{N}_{\ttheta\ttheta}(\ttheta_\text{pk})}e^{-\frac{(\pi-\theta_0)^2}{2q^2 N} - \frac{1}{2}} 
    = \frac{b}{2\sqrt{N\pi}\Cc_l^2}e^{-\frac{(\pi-\theta_0)^2}{2q^2 N} - \frac{1}{2}} \, ,
\end{equation}
where we approximated $\ttheta \approx \pi$ in the first term in the exponent in Eq.~\eqref{eq:Ccl_distribution_scaled} and used \eqref{eq:Ntilde_scaling} to substitute $\tilde{N}_{\ttheta\ttheta}(\ttheta_\text{pk}) = \tilde{N}_{\ttheta}^2(\ttheta_\text{pk})/b$.

The functional form of Eq.~\eqref{eq:Ccl_distribution_tail} is noteworthy: the distribution has a ``heavy'' tail that declines as $1/\Cc_l^2$, much slower than a Gaussian one. In fact, the decline is even slower than the exponential behaviour often encountered for the curvature perturbation $\zeta$ in PBH-producing single-field models (see e.g.,~\cite{Pattison:2017mbe, Ezquiaga:2019ftu, Figueroa:2020jkf, Biagetti:2021eep}; however, compare also with \cite{Hooshangi:2021ubn, Hooshangi:2023kss}, where the authors obtain sizeable tails for the statistics of  $\zeta$). We reproduce such an exponential behaviour for $\zeta$ for our model in Appendix~\ref{sec:curvature_distribution}. The difference highlights the importance of using $\Cc_l$ instead of $\zeta$ for the collapse criterion: in our benchmark cases below, we reach the threshold $\Cc_l=\Cc_{l,\text{th}}\sim 0.4$ for values of $\zeta$ that are smaller by one or more orders of magnitude. 

\begin{figure}
    \centering
    \includegraphics{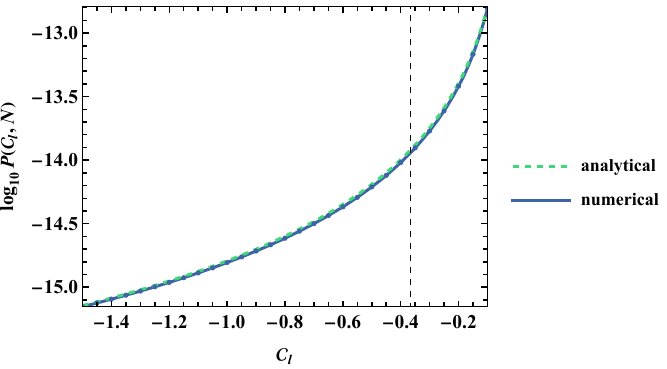}
    \caption{\it The tail of the probability distribution of $\Cc_l$ in point~A from Table~\ref{tab:benchmarks}. The numerical result is integrated numerically from Eq.~\eqref{eq:Ccl_distribution_scaled}. The analytical result is Eq.~\eqref{eq:Ccl_distribution_tail}.}
    \label{fig:Cltail}
\end{figure}

Fig.~\ref{fig:Cltail} compares the analytical approximation \eqref{eq:Ccl_distribution_tail} to the full result in a
specific example. For the full result, we first solve Eq.~\eqref{eq:theta_eom} numerically to obtain $\tilde{N}$ and then numerically integrate Eq.~\eqref{eq:Ccl_distribution_scaled}. The match between
 numerical and analytical results
is excellent. We notice the same behaviour  for all of the benchmark scenarios we are going to
discuss in Section \ref{sec:benchmark_points}.

To evaluate the PBH abundance, we must then integrate $P(\Cc_l, N)$ over the tail:
\begin{equation} \label{eq:beta_N}
    \beta_N = \int_{-\infty}^{\Cc_{l,\text{th}}} \dd \Cc_l \, P(\Cc_l,N) \, .
\end{equation}
To handle this equation, instead of starting from the result \eqref{eq:Ccl_distribution_tail}, we find it more convenient to begin again from Eq.~\eqref{eq:Ccl_distribution_scaled}. We perform the $\Cc_l$ integral first, yielding
\begin{equation} \label{eq:beta_in_erfc}
    \beta_N = \frac{1}{2 q \sqrt{2\pi N} } \int_{-\pi}^{\pi} \dd \ttheta \exp[-\frac{(\ttheta-\theta_0)^2}{2q^2N}] \text{erfc}\qty(\frac{|\Cc_{l,\text{th}}|}{\sqrt{2}q|\tilde{N}_{\ttheta}|}) \, .
\end{equation}
We restrict the domain of integration to $\ttheta \in [-\pi,\pi]$ and note that the integral is dominated by the  region $\ttheta \sim \pi$.\footnote{For $\ttheta > \pi$, the curvaton ends up on the other side of the hilltop and rolls to the minimum at $\ttheta=2\pi$, leading to the formation of a domain wall. We will briefly discuss such a scenario in Section~\ref{sec:discussion_and_conclusions}. Neglecting the domain wall dynamics, we may expect the integral \eqref{eq:beta_N} to receive similar contributions from the domains $\ttheta \lesssim \pi$ and $\ttheta \gtrsim \pi$ by symmetry (while the contributions away from $\ttheta \sim \pi$ are exponentially suppressed)---increasing the final result by a factor of $2$. Such an order one correction goes beyond the accuracy of our computation, and we omit it for simplicity.}
To evaluate the integral, we can further restrict the domain close to $\ttheta \approx \ttheta_\text{pk}$ from Eq.~\eqref{eq:saddle_point_approx}, in other words, to the region where the error function has significant support. We can then use formula \eqref{eq:Ntilde_scaling} to obtain
\begin{equation} \label{eq:beta_integrated}
    \beta_N \approx \frac{b}{2\pi\sqrt{N}|\Cc_{l,\text{th}}|} \exp[-\frac{(\pi-\theta_0)^2}{2q^2N}] \, ,
\end{equation}
where we again approximated $\ttheta\approx\pi$ in the exponent.\footnote{Integrating our previous result \eqref{eq:Ccl_distribution_tail} directly yields a similar result, but with a prefactor $e^{-1/2}/(2\sqrt{\pi})\approx0.17$ instead of $1/(2\pi)\approx0.16$.} The result depends on the model parameters $\theta_0$, $q$, $f_a$, and $R_\text{dec}$ through Eq.~\eqref{eq:b}---in particular, radiation dominating over the curvaton suppresses the perturbations and $\beta_N$ by making $R_\text{dec}$ large and $b$ small. This is a general effect for curvaton models, see, e.g., \cite{Bartolo:2003jx}. However, $\beta_N$ is most sensitive to the factor $q$ through the exponent. The result has a simple interpretation: the Gaussian factor in Eq.~\eqref{eq:beta_integrated} gives the probability for the field to drift to the hilltop, and the $b$ factor gives the width of the field interval around the hilltop.
The hilltop is responsible for producing the PBHs. Notice that this result does not directly depend on the ALP energy scale given by $m_a$ or $\Lambda_a$, though this scale determines the mass of the PBH, as we will see in Section~\ref{sec:PBH_numbers}.

\subsection{Comments on the approximations made}
\label{sec:comments_on_approximations}

Before moving on to discuss benchmark cases, we comment on the approximations made so far. Below
Eq.~\eqref{eq:compaction_function_in_zeta}, using the result \eqref{eq:theta_stochastics}, we approximated the linear compaction function as
\begin{equation} \label{eq:Cl_vs_dN}
    \Cc_l = -\frac{\dd \zeta}{\dd N} = -\tilde{N}_{\ttheta} \frac{\dd \ttheta}{\dd N}
    = -\frac{q\tilde{N}_{\ttheta}}{\sqrt{\dd N}}\xi_N \, .
\end{equation}
We then set $\dd N = 1$. This is a conventional choice---an e-fold is the typical scale of change during inflation---but, ultimately, $\dd N$ is a free parameter in our method. Setting $\dd N$ to some other value corresponds to replacing $\Cc_l$ with $\Cc_l\times \sqrt{\dd N}$ in all formulas such as Eq.~\eqref{eq:Ccl_distribution_scaled}, and the same for $\Cc_{l,\text{th}}$ in Eq.~\eqref{eq:beta_integrated}. A smaller $\dd N$ thus enhances the PBH abundance---it corresponds to changes over smaller distances, giving higher radial derivatives, leading to a higher $\Cc_l$.

There is a motivation for us to consider smaller $\dd N$ values. In Eq.~\eqref{eq:Cl_vs_dN}, we considered
 $\tilde{N}_{\ttheta}$  as approximately a constant over the relevant field range $\dd \ttheta \sim q\sqrt{\dd N}$. In reality, $\tilde{N}_{\ttheta}$ grows for perturbations that move $\ttheta$ towards $\pi$, enhancing PBH production---until the perturbations become large enough to take the field over the potential maximum and to the other side, where $|\tilde{N}_{\ttheta}|$ starts to rapidly decline (by symmetry). In order
 to avoid such complications, we need $\dd N \lesssim b/|\Cc_{l,\text{th}}|$.\footnote{This estimate arises from $\tilde{N}_{\ttheta\ttheta}(\ttheta_\text{pk})q\sqrt{\dd N} \lesssim \tilde{N}_{\ttheta}(\ttheta_\text{pk})$, or equivalently from $q\sqrt{\dd N} \lesssim \pi - \ttheta_\text{pk}$, where $q\sqrt{\dd N} \sim \dd \ttheta$ is the typical variation for the final kick that procudes $\Cc_l$ at $\ttheta=\ttheta_\text{pk}$.} In practice, this tends to be some orders of magnitude below the unit value, as we will see for our benchmark points below.

To resolve the dependence on $\dd N$, a more detailed analysis will be  needed, taking into account the full $\Cc$ profiles (as done for example in \cite{Raatikainen:2023bzk} for a single-field inflection-point model), but this is challenging and likely requires demanding numerics. We note, though, that our PBH abundance $\beta_N$ depends on $\dd N$ only polynomially---$\dd N$ does not affect the exponential factor in Eq.~\eqref{eq:beta_integrated}, which drives the PBH statistics. We thus set $\dd N$ to the conventional value of $1$, and assume the results to give an indicative---if not completely accurate---estimate of PBH production.

\section{Benchmark points for PBH production}
\label{sec:benchmark_points}


In this Section, focusing on the benchmark scenarios  summarized 
in Table~\ref{tab:benchmarks}, we discuss consequences of our previous results for the production of PBHs.


\subsection{PBH mass and abundance}
\label{sec:PBH_numbers}

As explained above, after inflation, a Hubble-sized region may collapse into a black hole if the energy density  is sufficiently large, as described by our collapse criterion $\Cc > \Cc_\text{th}$. The mass of the black hole approximately equals the total energy within the Hubble patch (ignoring order one corrections from matching the Fourier modes with real space lengths, see e.g., \cite{Young:2019osy}, and critical collapse, see e.g., \cite{Musco:2008hv, Luo:2020dlg})\footnote{For easier tracking of dimensions and magnitudes, we restore $\MPl$ to the equations in this section and in other observationally relevant results below.}:
\begin{equation} \label{eq:PBH_mass_from_energy_density}
    M_\text{PBH} = \frac{4\pi}{3}H^{-3}\rho
    = \frac{4\pi\MPl^2}{H} = \frac{4\sqrt{3}\pi\MPl^3}{\sqrt{\rho}} \, .
\end{equation}
Larger PBHs form at later times, when the Universe's energy density $\rho$ is smaller. The collapse starts when the perturbations' characteristic scale re-enters the Hubble radius. Different scales then correspond to different Hubble radii and different masses.
They follow the relation \cite{Tomberg:2024chk}
\begin{equation} \label{eq:M_PBH}
    M_\text{PBH} \approx \SI{1.11e49}{g} \times e^{-2(N-N_*)} \, .
\end{equation}
As above, $N$ here is the inflationary number of e-folds, $N_* \approx 5$ is the CMB pivot scale, and the corresponding wavenumber $k$ and comoving length $r$ are given in Eq.~\eqref{eq:N_vs_k_vs_r}.

After the PBHs  form, they behave like cold dark matter, and their energy density fraction grows relative to the surrounding radiation. Taking this property  into account, the PBH energy density fraction today is \cite{Tomberg:2024chk}
\begin{equation} \label{eq:Omega_PBH}
    \Omega_\text{PBH} = \num{4.09e14} \times \beta \qty(\frac{M_\text{PBH}}{\SI{e20}{g}})^{-1/2} \, ,
\end{equation}
where $\beta$ is the fraction of PBHs at formation.

PBHs  form at many different scales, over a range of masses. The PBH mass spectrum is conventionally described as
\begin{equation} \label{eq:f_PBH}
    f_\text{PBH} \equiv \frac{1}{\Omega_\text{DM}}\frac{\dd \Omega_\text{PBH}}{\dd \ln M_\text{PBH}} \, .
\end{equation}
Here $\Omega_\text{DM}$ is the total dark matter energy density today; we use the value  $\Omega_\text{DM} = 0.264$, taken from \cite{ParticleDataGroup:2022pth}. Note the slight difference in notation between \eqref{eq:f_PBH} and \eqref{eq:Omega_PBH}: what was called $\Omega_\text{PBH}$ in \eqref{eq:Omega_PBH} is called $\dd \Omega_\text{PBH}$ in \eqref{eq:f_PBH}, but it is the abundance coming from PBHs in a limited mass bin of width $\dd \ln M_\text{PBH}$. The total fraction of dark matter in the form of PBHs is then
\begin{equation} \label{eq:f_PBH_tot}
    f_\text{PBH,tot} \equiv \int \dd \ln M_\text{PBH} f_\text{PBH} \, .
\end{equation}

We notice here a subtlety related to the bin width $\dd \ln M_\text{PBH}$ in Eq.~\eqref{eq:Omega_PBH}. Below Eq.~\eqref{eq:compaction_function_in_zeta}, we considered changes in the curvature $\zeta$ over a step of $\dd N = 1$. We identify this step with the PBH bin width, so that $\beta_N$ in Eq.~\eqref{eq:beta_integrated} is the initial abundance of PBHs between scales $N-\frac{1}{2}$ and $N+\frac{1}{2}$. When computing $f_\text{PBH}$ with Eq.~\eqref{eq:f_PBH}, we replace $\dd \Omega_\text{PBH} \to \Omega_\text{PBH}$ from Eq.~\eqref{eq:Omega_PBH} and use there $\beta \to \beta_N$ from \eqref{eq:beta_integrated}; then, by Eq.~\eqref{eq:M_PBH}, we must also replace $\dd \ln M_\text{PBH} \to 2|\dd N| = 2$.\footnote{In Section~\ref{sec:comments_on_approximations}, we considered different step lengths $\dd N$. Decreasing the step also decreases the bin width, increasing $f_\text{PBH,tot}$. However, the formalism here assumes that the collapse probabilities from different bins are independent; for small $\dd N$, this assumption breaks down, presumably stabilizing $f_\text{PBH,tot}$.}

Both the $N$-dependence of $\beta_N$ in Eq.~\eqref{eq:beta_integrated}, and the explicit $M_{\rm PBH}$-dependence in Eq.~\eqref{eq:Omega_PBH},  lead to a PBH spectrum which increases towards smaller masses. However, the validity of the formalism in Section~\ref{sec:analytical_solutions} 
requires that PBHs form
  {\it after} the curvaton  decays  into radiation. Using
  relation \eqref{eq:PBH_mass_from_energy_density}, we find that the
   lowest  PBH mass
   we can consider is
\begin{equation} \label{eq:M_dec}
    M_\text{dec} \equiv \frac{4\sqrt{3}\pi\MPl^3}{R_\text{dec}^{1/2}\Lambda^2_a} \, .
\end{equation}
Since low masses dominate $f_\text{PBH}$, we take $M_\text{dec}$ to be the representative PBH scale for our benchmark points. 

For masses below $M_\text{dec}$, the curvaton perturbations are an isocurvature component, and they induce a time-dependent, growing curvature perturbation. The standard formalism for PBH formation from adiabatic perturbations does not directly apply; see \cite{Passaglia:2021jla,Domenech:2021and} for  studies of such a set-up. 
 We make a simplistic estimate of the PBH abundance at these scales, by computing the curvature perturbation and the 
value of $\Cc_l$ at the time the scales re-enter within the  Hubble radius. The formalism of Section~\ref{sec:analytical_solutions} can then be applied as if the curvaton decays at the 
moment of  re-entry. The parameter $R_\text{dec}$ becomes a function of $M_\text{PBH}$:
\begin{equation} \label{eq:Rdec_sliding}
    R_\text{dec} = \frac{48\pi^2 \MPl^6}{M_\text{PBH}^2\Lambda_a^4} \, .
\end{equation}
In Eq.~\eqref{eq:beta_integrated}, we then have $\beta_N \sim b \propto R_\text{dec}^{-1/4} \propto M_\text{PBH}^{1/2}$. This cancels the explicit $M_\text{PBH}$ dependence in Eq.~\eqref{eq:Omega_PBH}---we are converting the oscillating curvaton perturbations into the form of PBHs, but since both quantities scale as cold dark matter, the timing of the conversion makes no difference. There is still an implicit scale dependence
in our formulas through the quantity $N$ in Eq.~\eqref{eq:beta_integrated}---but it is milder than in the post-decay case.

The isocurvature regime does not extend to arbitrarily small PBH masses. Indeed, when $\rho \gg \Lambda^4$, the ALP potential does not exist, and neither do the curvature perturbations. The corresponding mass cutoff is
\begin{equation} \label{eq:M_Lambda}
    M_\Lambda \equiv \frac{4\sqrt{3}\pi\Mpl^3}{\Lambda_a^2} \, .
\end{equation}
Another cutoff scale is given by the time the curvaton starts to oscillate: before this epoch, the curvaton is frozen and its energy density is constant. The latter is  more and more subdominant with respect to the radiation bath as we move backwards in time, leading to a negligible curvature perturbation. In Section~\ref{sec:analytics_away_from_hilltop}, we learned that the curvaton  oscillations typically start when $\rho \sim m_a^2\Mpl^2$, giving the mass cutoff\footnote{Below, we will see that reasonable models have $f_a < \MPl$. From this, we see that $M_\Lambda/M_\text{osc} = \MPl m_a/\Lambda_a^2 = \MPl/f_a > 1$. In priciple, this means that the curvaton starts to oscillate before the potential has reached its final form, and we should use a temperature-dependent potential with a mass given by Eq.~\eqref{eq:mofT}. In practice, this temperature dependence increases the factor of $n$ in the exponents of Eqs.~\eqref{eq:phi_approx_eom_2} and \eqref{eq:small_theta_approx_eom} at early times. This modifies $n_\text{osc}$ slightly, but we do not expect the modification to change the qualitative properties of the model.}
\begin{equation} \label{eq:M_osc}
    M_\text{osc} \equiv \frac{4\sqrt{3}\pi\Mpl^2}{m_a} \, .
\end{equation}

%



\begin{table}
\begin{center}
\begin{tblr}{colspec={
    Q[halign=l]
    Q[halign=c]
    Q[halign=c]
    Q[halign=c]
  }}
\toprule
& \SetCell[c=3]{c} \textbf{Benchmark points} \\
& A & B & C \\
\midrule
\textbf{Input parameters} &&& \\
$M_\text{PBH}$ & \SI{e21}{g} & \SI{e19}{g} & \SI{e29}{g} \\
$f_\text{PBH}$ & 1 & 1 & 0.005 \\
$\Pp_{\zeta,\psi}(k_*)$ & \num{e-10} & \num{e-15} & \num{e-12} \\
$H_*/H_\text{max}$ & 1 & 1 & 0.1 \\
$\theta_0$ & 0.2 & 1.5 & 0.5  \\
\midrule
\textbf{Model parameters} &&& \\
$f_a$ & \num{4.6e-5}$\Mpl$ & \num{7.6e-5}$\Mpl$ & \num{3.8e-6}$\Mpl$ \\
$m_a$ & \SI{5.4e14}{eV} & \SI{1.6e9}{eV} & \SI{1.2e8}{eV}  \\
$q$ & 0.0667 & 0.0403 & 0.0809 \\
$\Lambda_a$ & \SI{7.8e9}{GeV} & \SI{1.7e7}{GeV} & \SI{1.0e6}{GeV}  \\
$R_\text{dec}$ & \num{8.6e-17} & 0.036 & \num{2.6e-17} \\
\midrule
\textbf{PBH distribution} &&& \\
$\beta_N$ & \num{4.1e-15} & \num{4.1e-16} & \num{2.0e-13} \\
$f_\text{PBH,tot}$ & 0.59 & 0.66 & \num{3.0e-3} \\
$M_\Lambda$ & \SI{9.3e12}{g} & \SI{1.9e18}{g} & \SI{5.1e20}{g} \\
\midrule
\textbf{Inflation} &&& \\
$N_\text{PBH}$ & 37.7 & 40.0 & 28.4 \\
$r$ & 0.036 & 0.036 & \num{3.6e-4} \\
\midrule
\textbf{Hilltop characteristics} &&& \\
$b$ & \num{9.6e-3} & \num{6.1e-6} & \num{3.5e-4} \\
$\pi - \ttheta_\text{pk}$ & \num{1.7e-3} & \num{6.7e-7} & \num{7.8e-5} \\
$\zeta(\ttheta_\text{pk}) = \tilde{N}(\ttheta_\text{pk}) - \tilde{N}(\theta_0)$ & 0.054 & \num{7.3e-5} & \num{2.9e-3} \\
\bottomrule
\end{tblr}
\end{center}
\caption{\it Parameter values for benchmark points. Points A and B give PBH dark matter in the asteroid mass window and are optimized for large $b$ and large $R_\text{dec}$, respectively. Point C produces subdominant dark matter of heavier black holes.}
\label{tab:benchmarks}
\end{table}

\subsection{Fitting the parameters}
\label{sec:fitting_parameters}

We are now in the position to make and motivate a choice of benchmark points for
interesting scenarios leading to curvaton PBH production.  We describe  
the criteria we follow to make our selection.

To build our benchmark points, we choose values for the following quantities:
the PBH mass scale $M_\text{PBH}$;
the PBH dark matter fraction $f_\text{PBH}$, as obtained from this mass scale;
the inflationary Hubble parameter $H_*$ (which should satisfy Eq.~\eqref{eq:CMB_constraints});
the curvaton $\psi$-contribution to the CMB power spectrum $\Pp_{\zeta,\psi}(k_*)$ (which should be subdominant to the measured amplitude  of $2.1\times 10^{-9}$); and
finally, the initial field value $\theta_0$.
    
Choosing these
quantities uniquely fixes all the model parameters,
through the following considerations:
The PBH mass corresponds to a given inflationary e-fold number $N$, and with $f_\text{PBH}$ and Eqs.~\eqref{eq:Omega_PBH}, \eqref{eq:f_PBH}, this quantity  also sets the initial PBH fraction $\beta$.
     The values of $\Pp_{\zeta,\psi}$ and $\theta_0$ fix the parameter combination $qf_a^{3/2}/R_\text{dec}^{1/4}$ by means of Eqs.~\eqref{eq:power_spectrum_theta} and \eqref{eq:Ntilde_approx_3}.
    Substituting $\beta$, $qf_a^{3/2}/R_\text{dec}^{1/4}$, $\theta_0$, and $N$ into Eq.~\eqref{eq:beta_integrated} with Eq.~\eqref{eq:b} gives us a condition that we can use to obtain $q$.
     The definition of $q$ in Eq.~\eqref{eq:scaled_variables}, together with $H_*$, gives us
    the value of $f_a$, and using this quantity and the value for $qf_a^{3/2}/R_\text{dec}^{1/4}$, we obtain $R_\text{dec}$.
     Setting $M_\text{PBH}=M_\text{dec}$ in Eq.~\eqref{eq:M_dec} we can solve for $\Lambda_a$ and, hence, for $m_a$.

 
 Some additional comments are in order, with respect to our choice of benchmark points. 
 The PBH fraction $f_\text{PBH}$ is predominantly determined by the exponent in Eq.~\eqref{eq:beta_integrated}, which has to be of order ten for ensuring $f_\text{PBH} \sim 1$. For typical values of $\theta_0 - \pi \sim 1$, $N \sim 10$, this conditions sets $q\sim 0.01\dots 0.1$. Larger $q$ values lead to PBH overproduction, while lower $q$ values produce a negligible abundance of PBHs.
The value of $\theta_0$ does not affect the results excessively, but it has to be of order one in
order to produce reasonable values for $\Pp_{\zeta,\psi}$.  $\theta_0$  can not be too close to the hilltop at $\theta=\pi$; on the other hand, if it is too close to zero, the approximation \eqref{eq:power_spectrum_theta}  breaks down, and we need to use the results of Section~\ref{sec:analytics_away_from_hilltop},
leading to an effective value of order $\theta_0 \sim q\sqrt{N} \sim 0.1$ (the last approximation is based on the  considerations above for $q$ and $N$). Benchmark point A is chosen to saturate this limit.

With $q$ fixed and $\theta_0 \sim 1$, the power spectrum $\Pp_{\zeta,\psi}$ depends mainly on the combination $f_a^{3/2}/R_\text{dec}^{1/4}$---essentially, the quantity  $b$ from Eq.~\eqref{eq:b}. To suppress $\Pp_{\zeta,\psi}$, $b$ has to be small. By the arguments of Section~\ref{sec:comments_on_approximations}, this is not what we wish: we would like to have $b$ be at least of order $0.1$ for our computation to be reliable with a bin width of $\dd N \sim 1$. Unfortunately, we can not  reach such a value. The example point $A$ is tuned to produce a large $b$ by increasing $\Pp_{\zeta,\psi}$  and decreasing $\theta_0$.
Still, 
we  only get $b \sim 0.01$.
A large $b$ implies either a low $R_\text{dec}$ or a high $f_a$. Note that a low $R_\text{dec}$ implies a large separation between the mass scales $M_\text{dec}$ and $M_\Lambda$. As discussed above, this region is not well described by our computation, and we would like to minimize it---rather than lower $R_\text{dec}$, we would like to increase $f_a$. However, for $q \gtrsim 0.01$, the $H_*$ bound \eqref{eq:CMB_constraints} gives an upper bound for $f_a$:
\begin{equation} \label{eq:f_a_max}
    f_a \lesssim \frac{H_\text{max}}{2\pi \times 0.01} = \num{3.1e-4}\MPl = \SI{1.3e15}{GeV} \, .
\end{equation}
To make the quantity $f_a$  as large as possible, we  choose $H_* = H_\text{max}$ for benchmark points A and B. For point A, the resulting $R_\text{dec} \sim 10^ {-17}$ is still considerably low. The only way to rise $R_\text{dec}$ would be to decrease the value of  $\Pp_{\zeta,\psi}$, and accept a small value for $b$. Benchmark point B is optimized to do this.

Furthermore,
we also make a number of consistency checks that the benchmark points satisfy. We demand that:
The Peccei--Quinn symmetry gets broken before the onset of inflation,  that is, $q<1$;
the scale $f_a$ is sub-Planckian;
the linear approximation for the power spectrum is reliable at the CMB scales, that is, $\tilde{N}$ does
      not vary much at those scales: $|\tilde{N}_{\ttheta\ttheta}(\theta_0)|q\sqrt{N_*} \lesssim |\tilde{N}_{\ttheta}(\theta_0)|$;
the process of PBH formation starts below the inflationary Hubble scale: $3H_*^2\MPl^2 > \Lambda_a^4$;
the PBH-forming scalar field perturbations from $\ttheta_\text{pk}$ already start  to oscillate by the epoch of scalar decay: $n_\text{osc}(\ttheta_\text{pk}) < -\frac{1}{2}\ln f_a - \frac{1}{2}\ln R_\text{dec}$ (using the definition of $n$);
and the scalar is still subdominant compared to the background radiation at the decay time: $4\tilde{N}(\ttheta_\text{pk}) < \ln 2$ (from Eq.~\eqref{eq:decay_N_eq_2}).
Most of these conditions are satisfied automatically as per our discussion above.

\begin{figure}
    \centering
    \includegraphics{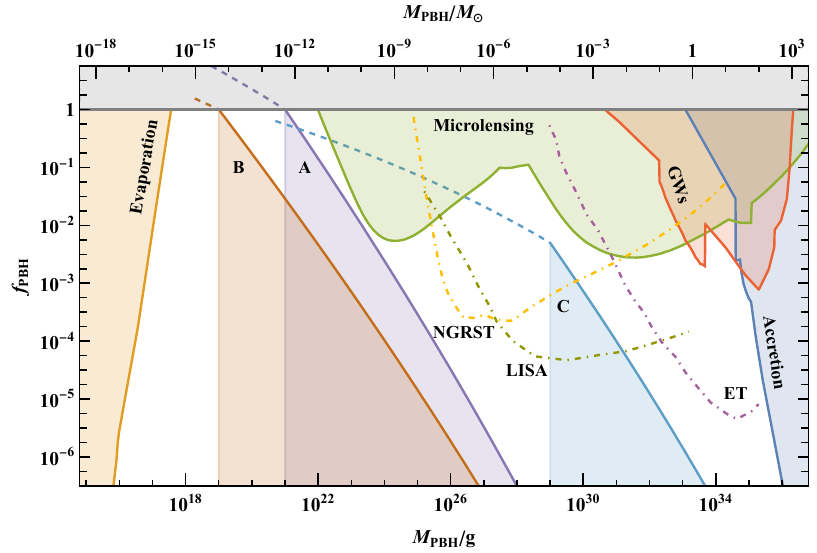}
    \caption{\it Benchmark mass distributions $f_\text{PBH}$ plotted against dominant PBH bounds. The upper shaded regions are excluded by PBH evaporation, microlensing observations, gravitational wave constraints, and accretion limits. The constraints are adapted from Figure~4 of \cite{Green:2024bam}; the data is available at \cite{KavanaughPBHbounds}. The projected sensitivities of the Nancy Grace Roman Space Telescope (NGRST) \cite{DeRocco:2023gde}, LISA \cite{DeLuca:2021hde,Pujolas:2021yaw,Franciolini:2022htd,Martinelli:2022elq,Franciolini:2023opt}, and Einstein Telescope (ET) \cite{Branchesi:2023mws} are plotted as dot-dashed lines, see Figure~5 in \cite{DeRocco:2023gde}. The constraints are for monochromatic PBHs, and the comparison to our mass functions is approximative; for details, see \cite{Carr:2017jsz}.}
    \label{fig:mass_distributions}
\end{figure}

\subsection{PBH mass spectra}
\label{sec:mass_spectra}

We now discuss the PBH mass spectra obtained in our scenario.
Figure~\ref{fig:mass_distributions} shows the mass distributions $f_\text{PBH}(M)$ of our benchmark points. The solid lines correspond to the reliable region of masses above $M_\text{dec}$, while the dashed lines are the continuations into smaller masses, down to $M_\Lambda$. Since the standard formalism does not apply in such small-mass regime, the dashed lines are uncertain. Furthermore, since they are characterized by  higher values of $R_\text{dec}$, they correspond to lower values of $b$, and suffer also from the low-$b$ problems described in Section~\ref{sec:comments_on_approximations}. Realistic PBH abundance may be suppressed at these small-mass scales.
 For these reasons, we neglect such mass scales in this work. In particular, we compute the total PBH energy density fraction $f_\text{PBH,tot}$ in Eq.~\eqref{eq:f_PBH_tot} only considering PBH masses larger than our reference scale $M_\text{PBH}$, integrating over the shaded regions in Figure~\ref{fig:mass_distributions}. We notice that the shaded portions of the spectra in Figure~\ref{fig:mass_distributions} satisfy current observational constraints; points A and B produce all dark matter in the asteroid mass window, while point C gives a subleading dark matter population at larger masses, testable by upcoming GW and gravitational lensing experiments.

\subsection{Estimate of fine-tuning}
\label{sec:fine_tuning}
Single-field PBH models of inflation are usually highly fine-tuned \cite{Cole:2023wyx} (though a double-well potential or a potential with a step-like feature may alleviate this somewhat \cite{Karam:2023haj, Stamou:2024lqf}). As already emphasized in \cite{Stamou:2023vft}, a spectator field model like ours may avoid such fine-tuning. Following \cite{Cole:2023wyx, Azhar:2018lzd}, we quantify the level of fine-tuning in an observable $\mathcal{O}$ with respect to a parameter $p$ as
\begin{equation} \label{eq:fine_tuning_epsilon}
    \epsilon_\mathcal{O} \equiv \frac{\dd \ln \mathcal{O}}{\dd \ln p} \, .
\end{equation}
Small values of $\epsilon_\mathcal{O}$ correspond to less fine-tuning.
We are interested in the observable $f_\text{PBH}$, computed via Eqs.~\eqref{eq:f_PBH} and \eqref{eq:beta_integrated}. It depends on most model parameters in a power law-like manner, $f_\text{PBH} \sim p^n$, giving $\epsilon_{f_\text{PBH}} = n$, where $n$ is of order one. There is no considerable fine-tuning with respect to these parameters. Interestingly, the collapse threshold $\Cc_{l,\text{th}}$ is among them.

The black hole fraction depends more strongly on the parameters in the exponent in Eq.~\eqref{eq:beta_integrated}, namely, $\theta_0$ and, in particular, $q$, which we already noted to be the determining factor in PBH formation.\footnote{Since $\theta_0$ is an initial condition rather than a fundamental model parameter, one may argue its value is set anthropically, but this does not remove the issue of fine-tuning \cite{Stamou:2023vft}.} We have
\begin{equation} \label{eq:epsilon_values}
     q\!: \ \epsilon_{f_\text{PBH}} = \frac{(\pi-\theta_0)^2}{q^2N} \, , \qquad \theta_0\!: \ \epsilon_{f_\text{PBH}} = \frac{(\pi-\theta_0)\theta_0}{q^2N} \, .
\end{equation}
These are both still comparable to the exponent in $\beta$, which in turn must be approximately of order ten to not suppress the PBH abundance too much. The exact numbers for our benchmark points are given in Table~\ref{tab:finetuning}. These numbers show that the model is not very fine-tuned, especially compared to the  values of $10^4\dots 10^9$ discussed in \cite{Cole:2023wyx} for the case of single-field inflationary scenarios. Indeed, the PBH abundance does not go through the usual mechanism of a highly-tuned enhanced curvature power spectrum; instead, we only need to guarantee that the quantum diffusion scale $q$ is comparable to the field shift $\pi - \theta_0$ needed to reach the potential hilltop.

\begin{table}
\begin{center}
\begin{tblr}{colspec={
    Q[2.0cm,halign=r]
    Q[1.8cm,halign=c]
    Q[1.8cm,halign=c]
    Q[1.8cm,halign=c]
  }}
\toprule
& A & B & C \\
\midrule
$q\!: \ \epsilon_{f_\text{PBH}}$ & 51.7 & 41.5 & 37.5 \\
$\theta_0\!: \ \epsilon_{f_\text{PBH}}$ & 3.5 & 37.9 & 7.1 \\
\bottomrule
\end{tblr}
\end{center}
\caption{\it Fine-tuning for the benchmark points of Table~\ref{tab:benchmarks}.}
\label{tab:finetuning}
\end{table}
\section{Extensions of the analysis}
\label{sec:applications}
\label{sec:particle_dark_matter_scenario}

While our work so far focussed on PBH production in the axion-like curvaton model, it can also find interesting applications in related areas. We discuss them in this Section. 

\subsection{Mixed PBH and ALP dark matter}



Until now,   we have assumed that the spectator scalar field  decays into radiation after inflation ends, converting
  its fluctuations into adiabatic curvature perturbations
  as in a standard curvaton setup. However, we can also consider a case where the conversion
  is not complete, and the ALP survives to these days, constituting (a fraction of) dark matter.
   See e.g. \cite{Marsh:2015xka, DiLuzio:2020wdo} for reviews of this possibility\footnote{We only consider here a leftover ALP condensate arising from the misalignment mechanism, omitting ALP production from other channels, such as the decay of topological defects and interactions with other particles \cite{Marsh:2015xka}.}. Let us briefly
   discuss this case.

First of all, 
if the ALP field does not completely decay into radiation, we need to be mindful of the  isocurvature perturbation 
it induces  \cite{Tenkanen:2019aij, Langlois:2004nn},
\begin{equation} \label{eq:isocurvature_pert}
    S = -\frac{\delta\rho_\psi}{\bar{\rho}_\psi} \, ,
\end{equation}
where $\rho_\psi$ is the scalar field energy density,  $\delta\rho_\psi$  is its perturbation, and $\bar{\rho}_\psi$ is its mean. The quantity 
$S$ stays constant at super-Hubble scales, until the scalar field starts to oscillate.
It corresponds, up to an order one factor, to the contribution of the scalar field to the curvature power spectrum,
 in a phase when the ALP dominates the energy density. Hence, it is important during recombination, when matter density is comparable to radiation density.

To compute the value of $S$ in Eq.~\eqref{eq:isocurvature_pert}, we use the formalism of Section~\ref{sec:analytical_solutions}. We write $\rho_\psi$ at time $N$, assuming the rescaled field is frozen at $\ttheta$ during inflation (compare with Eq.~\eqref{eq:decay_N_eq}):
\begin{equation} \label{eq:rho_theta_perts}
\begin{gathered}
    \rho_\psi(\ttheta,N_p) = g(\ttheta)\Lambda_a^4e^{-3[N_p-N_{p,\text{osc}}(\ttheta)]} \\
    \Rightarrow \frac{\delta \rho_\psi}{\bar{\rho}_\psi}
    \approx \frac{\rho_\psi(\ttheta,N_p) - \rho_\psi(\theta_0,N_p)}{\rho_\psi(\theta_0,N_p)}
    \approx \qty[\frac{g'(\theta_0)}{g(\theta_0)} + 3N'_{p,\text{osc}}(\theta_0)]\delta\ttheta \, ,
\end{gathered}
\end{equation}
where we expanded $\ttheta = \theta_0 + \delta\ttheta$ to leading order around the mean value $\theta_0$. This condition gives the isocurvature power spectrum
\begin{equation} \label{isocurvature_powr_spectrum}
    \Pp_S = \qty|\frac{g'(\theta_0)}{g(\theta_0)} + 3N'_{p,\text{osc}}(\theta_0)|^2 \Pp_{\delta\theta}
    = \qty|\frac{g'(\theta_0)}{g(\theta_0)} + 3N'_{p,\text{osc}}(\theta_0)|^2 q^2 \, .
\end{equation}
At low $\theta_0$, we have $g(\theta_0) \approx \frac{1}{2}\theta_0^2$, so the first term in the sum is $2/\theta_0$, which grows as $\theta_0 $ goes to small values. As $\theta_0 \to \pi$, instead,  the second term starts to grow, as discussed in Section~\ref{sec:analytics_near_hilltop}. The combination appearing in Eq.~\eqref{isocurvature_powr_spectrum} then reaches a minimum in the middle region when $\theta_0$ is of order one. We thus obtain $\Pp_S \gtrsim q^2$. In scenarios in which the oscillating scalar field constitutes {\it all} of  the dark matter,  CMB constraints require \cite{Planck:2018jri, DiLuzio:2020wdo}
\begin{equation} \label{eq:P_S_bound}
    \Pp_S < 0.040 \, \Pp_R(k_*) \approx \num{8.3e-11} \, ,
\end{equation}
which translates into the condition $q \lesssim \num{e-5}$.

Let us then consider PBH production.
As discussed in Section~\ref{sec:PBH_numbers}, our PBH formalism does not apply when the scalar has not decayed, the scenario we are considering here. However, we can use the estimate outlined around Eq.~\eqref{eq:Rdec_sliding}, making $R_\text{dec}$ a function of the scale under consideration. We then recover our previous limit $q \gtrsim 0.01$ for significant PBH production. Comparing this to the above limit of $q \lesssim \num{e-5}$, this implies that  \emph{the ALP field scenario can not simultaneously provide significant fractions of both particle dark matter and PBH dark matter.} The same conclusion holds  for other scalar field models with similar potentials, see the analysis in
Appendix~\ref{sec:general_hilltop_computation}.

If the ALP only constitutes a small fraction of the dark matter, its contribution to effective isocurvature perturbations is weighted by this fraction squared. Setting $q=0.01$ for significant PBH production\footnote{Note that even though the scalar field causes the PBH formation, the PBH abundance can be much higher than the scalar field DM abundance, since the PBH mass consists mostly of visible matter, which dominates during the PBH collapse.} and using the bound \eqref{eq:P_S_bound}, we obtain the following bound for this fraction:
\begin{equation} \label{eq:Omega_psi_bound_with_PBHs}
    \frac{\Omega_\psi}{\Omega_\text{DM}} \lesssim 10^{-3} \, .
\end{equation}
In this formula,
the quantity $\Omega_\psi$ can be expressed  as
\begin{equation} \label{eq:Omega_psi}
    \Omega_\psi = \frac{\rho_{\psi,\text{osc}}}{3H_0^2\MPl^2}\times\qty(\frac{g_{s0}T_0}{g_s T_\text{osc}})^3
    = 7.0 \times 10^{-7} \qty(\frac{M_\Lambda}{\text{g}})^{-1/2} \qty(\frac{f_a}{\text{GeV}})^{3/2}\theta_0^2 \, .
\end{equation}
The quantity
 $\rho_{\psi,\text{osc}} = V(\theta_0) \approx \frac{\Lambda_a^4}{2}\theta_0^2$ is the scalar field energy density when it starts to oscillate, and the associated radiation temperature $T_\text{osc}$ is
 obtained from the relation $\rho_{r,\text{osc}} \approx m_a^2 \MPl^2 = \frac{g\pi^2}{30}T_\text{osc}^4$ (see our discussion below Eq.~\eqref{sec:n_osc_small_theta_0}).
 
  We use $T_0 \approx 2.7K$, $H_0 \approx \SI{70}{km/s/Mpc}$, $g_s=g=106.75$ (the  Standard Model degrees of freedom), and $g_{s0}=3.909$ (the entropy degrees of freedom today). We also express the model parameters in terms of the mass $M_\Lambda$ from Eq.~\eqref{eq:M_Lambda}, which we take to be the characteristic (minimum) PBH scale in the absence of decay.
Taking $\theta_0 \sim 1$ and $\Omega_\text{DM} = 0.264$, Eqs.~\eqref{eq:Omega_psi_bound_with_PBHs} and \eqref{eq:Omega_psi} give constraints for a curvaton PBH scenario
with
 a leftover ALP field.
The constraints can be expressed in many ways:
\begin{equation} \label{eq:DM_PBH_bounds}
\begin{gathered}
    M_\text{PBH} > M_\Lambda \gtrsim \SI{7e-6}{g} \times \qty(\frac{f_a}{\text{GeV}})^3 \, , \\
    M_\text{PBH} > M_\Lambda \gtrsim \SI{e30}{g} \times \qty(\frac{m_a}{\text{eV}})^{-3/4} \, , \\
    m_a \lesssim \SI{8e46}{eV}\times\qty(\frac{f_a}{\text{GeV}})^{-4} \, .
\end{gathered}
\end{equation}
We expressed the constraints in forms that are easy to compare to experimental axion searches, see e.g. Ref.~\cite{Machado:2019xuc, DiLuzio:2020wdo} for reviews. 
We can do the same for $M_\Lambda$, and, for completeness, for $M_\text{osc}$ from Eq.~\eqref{eq:M_osc}:
\begin{equation} \label{eq:mass_relations}
    M_\Lambda = \SI{5.6e41}{g} \times \qty(\frac{m_a}{\unit{eV}})^{-1} \qty(\frac{f_a}{\unit{GeV}})^{-1} \, , \quad
    M_\text{osc} = \SI{2.3e23}{g} \times \qty(\frac{m_a}{\unit{eV}})^{-1} \, .
\end{equation}
Taken together with the constrain $f_a \lesssim \SI{1.3e-15}{GeV}$ given in Eqs.~\eqref{eq:f_a_max}, \eqref{eq:Omega_psi_bound_with_PBHs} and \eqref{eq:DM_PBH_bounds} characterize the parameter space for significant PBH production in a scenario where the ALP does not decay. Note, in particular, that for the QCD axion, $m_a^2 f_a^2 = (\SI{75}{MeV})^4$ \cite{DiLuzio:2020wdo}, giving $M_\Lambda = 49 M_\odot$. This coincidence between the QCD scale and solar mass PBHs has also been noted before---the QCD phase transition softens the equation of state momentarily, enhancing PBH formation at these scales in models with a wide perturbation spectrum, see e.g.~\cite{Jedamzik:1998hc, Byrnes:2018clq}. Our setup offers another way to capitalize on this relationship to produce solar mass PBHs. 

\subsection{Higgs-like models}
\label{sec:higgs_PBHs}
Besides the ALP-scenario, we may consider other curvaton setups with a local maximum in their potential. A well-motivated example are Higgs-like models with a potential of the form
\begin{equation} \label{eq:higgs_like_potential}
    V(\psi) = \frac{\lambda}{4}\qty(\psi^2 - v^2)^2 \, .
\end{equation}
Just like the ALP potential, Eq.~\eqref{eq:higgs_like_potential} has a quadratic maximum (at $\psi = 0$) close to a quadratic minimum (at $\psi = \pm v$). We assume the field freezes in-between during inflation. After inflation, the field starts to oscillate around the minimum. PBHs form from patches where the field froze near the maximum. See also \cite{Fumagalli:2019ohr} for an analysis of stochastic spectactor
fields in the context of Higgs models.

The analytical computation of the PBH fraction $\beta$ is then analogous to the ALP case, with the identifications
\begin{equation} \label{eq:axion_higgs_correspondence}
    f_a \sim v \, , \quad m_a \sim \lambda v^2 \, , \quad \Lambda_a^4 \sim \lambda v^4 \, .
\end{equation}
Due to slight differences in the potentials, there are order one corrections to some of the formulas of Section~\ref{sec:analytical_solutions}. We present the details in Appendix~\ref{sec:general_hilltop_computation}, where we work out the general case of a curvaton with a quadratic hilltop. The results are qualitatively similar to the ALP case; for example, we can reproduce close equivalents of the benchmark points of Table~\ref{tab:benchmarks}. For strong PBH formation, we still need $q \sim H_*/v \sim 0.01$. This implies either low-scale inflation or a high value of $v$. The value of $\lambda$ can then be used to set the PBH mass scale, and must typically be small, as we will see below.

In the Higgs-like case, we need to be careful about the lower cutoff of the PBH mass. For the ALP, we argued this is set by the potential scale $\Lambda$, since at higher energies, the ALP potential is flat. Arranging such a mechanism in the Higgs-like case is not straightforward. However, we can instead take the lower cutoff to be $M_\text{osc}$ from Eq.~\eqref{eq:M_osc}, the scale where $3H^2 \approx m^2=2\lambda v^2$ and the field starts to oscillate, since before this, all field values are frozen and their contribution to the Universe's evolution is small and almost independent of the field value, suppressing the $\zeta$ derivative in Eq.~\eqref{eq:compaction_function_in_zeta}. For large $v$, the difference between the two scales is not big. For the Higgs, we have
\begin{equation} \label{eq:higgs_mass_relations}
\begin{aligned}
    M_\text{osc} &= \SI{6.7e20}{g} \times \qty(\frac{\lambda}{10^{-50}})^{-1/2}\qty(\frac{v}{\MPl})^{-1} \, , \\
    M_\Lambda &= \SI{3.8e21}{g} \times \qty(\frac{\lambda}{10^{-50}})^{-1/2}\qty(\frac{v}{\MPl})^{-2}\, .
\end{aligned}
\end{equation}
It is noteworthy, however, that if the field is coupled to the surrounding thermal bath, the potential may obtain thermal corrections. In particular, a thermal mass $m_T \sim T^2$ is always larger than $H^2 \sim T^2\times(T/\MPl)^2$, implying oscillations from the very beginning. This not only removes the PBH mass cutoff, but also drowns the hilltop in the potential, making the scenario unviable. To avoid this problem, the field must either be decoupled from the thermal bath before it decays, or it must be protected from thermal corrections by some other mechanism.

\subsection{PBH Clustering}
\label{sec:clustering}


We now discuss the effect of clustering in our
PBH scenario. 
Let us consider two patches of space with sizes $N_1$ and $N_2$, separated by a different, longer scale $N_c$. We denote the length and mass scales related to the PBHs by $r_1$, $r_2$, $M_1$, and $M_2$, and the length scale related to $N_c$ by $r$; the conversions are done by Eqs.~\eqref{eq:N_vs_k_vs_r} and \eqref{eq:M_PBH}. We would like to answer the following question: if the first patch collapses into a black hole, what is the probability that the second patch also collapses (denoted below by $p_c$)? This gives a measure of the initial black hole clustering, a phenomenon important for early structure formation, PBH microlensing constraints, and black hole merger rate. The merger rate further affects the final PBH mass distribution and the gravitational wave signal. For a review of the effects of PBH clustering, see \cite{Belotsky:2018wph}.

To answer the question, we note that the final, small patches of space  were part of the same larger patch for the first $N_c$ e-folds. Their stochastic evolution is identical in the beginning, and they develop independently only after $N_c$ e-folds. For each stochastic path that forms a PBH in the first patch, we need to sum the probabilities of all paths that deviate from the original path at $N_c$ but still form a PBH in the second patch.

The probability $P(\theta,N)$ to find the field at value $\theta$ at time $N$ is given by Eq.~\eqref{eq:theta_stochastics}. Generalizing Eq.~\eqref{eq:beta_integrated}, the probability to form a PBH at scale $N_b$ when starting from $\theta$ at $N_a$ (where $N_a < N_b$) is
\begin{equation} \label{eq:beta_from_Nc}
    \beta(N_b;N_a,\theta)
    = \frac{b}{2\pi\sqrt{N_b-N_a}|\Cc_{l,\text{th}}|} \exp[-\frac{(\pi-\theta)^2}{2q^2(N_b-N_a)}] \, .
\end{equation}
Then, by basic rules of probability, the conditional probability we are looking for is
\begin{equation} \label{eq:p_of_second_PBH}
\begin{aligned}
    p_c(N_1,N_2;N_c)
    &= \frac{\int_{-\infty}^\infty \dd \theta_c \, P(\theta_c,N_c)\beta(N_1;N_c,\theta_c)\beta(N_2;N_c,\theta_c)}{\beta(N_1;0,\theta_0)} \,, \\
    &= \frac{b}{2\pi|\Cc_{l,\text{th}}|}\sqrt{\frac{N_1}{N_1 N_2 - N_c^2}}e^{-\frac{(\pi-\theta_0)^2}{2q^2}\frac{(N_1 - N_c)^2}{N_1(N_1 N_2 - N_c^2)}} \, .
\end{aligned}
\end{equation}
The first $\beta$ factor gives the probability of a PBH forming in the first patch; it acts as the weight for the second $\beta$ factor, the probability of another PBH forming in the second patch. The factor in the denominator ensures the correct normalisation. In particular, when $N_c \to 0$, we have $P(\theta_c,N_c) \to \delta(\theta_0 - \theta_c)$, and
\begin{equation} \label{eq:p_of_second_PBH_limit_1}
    p_c(N;0) = \beta(N_2;0,\theta_0) = \beta_{N_2} \, , 
\end{equation}
that is, the result from Eq.~\eqref{eq:beta_integrated}---PBH formation in the second patch is at the background (Poissonian) level, independent of the first patch, which is far away.

In the opposite limit, $p_c$ grows as $N_c$ approaches the scales $N_1$ and $N_2$. The excess near $N_c \sim N_1, N_2$ indicates clustering: there is an increased probability to find PBHs next to each other. If $N_1 = N_2$, the result \eqref{eq:p_of_second_PBH} diverges in this limit; for $N_1 \neq N_2$, $p_c \to 0$ as $N_c \to \sqrt{N_1 N_2}$, but it exhibits a large peak before plummeting to zero. The peak may exceed unit value, which
is not allowed by basic properties of
 a probability distribution (see \cite{Ali-Haimoud:2018dau} for a similar argument). This happens because our approximation for $\beta(N_b;N_a,\theta)$ breaks down when $N_b \to N_a$: Eq.~\eqref{eq:beta_from_Nc} assumes that $\theta$ is far from $\pi$ and that the stochastic motion takes it there only after the $N_b-N_a$ e-folds. Hence, Eq.~\eqref{eq:p_of_second_PBH} becomes unreliable in this limit. Notice also that we have neglected the cloud-in-cloud problem: the PBHs can not form so close to each other that they would overlap. For these reasons, we force $p_c \leq 1$ by hand, and we also set $p_c(N_1,N_2;N_c) = 1$ for $N_c \geq N_1$ or $N_c \geq N_2$.\footnote{Often (see e.g.~\cite{Animali:2024jiz, Auclair:2024jwj}), $p_c$ is taken to be zero at small separations with the interpretation that a PBH can not form inside another PBH. With this convention, the corresponding quantity $\xi_\text{PBH}$ in Eq.~\eqref{eq:clustering_xi} equals $-1$ in this limit. We adopt $p_c=1$ in this regime instead, since it matches the growing behaviour of $p_c$. In our interpretation, the probability to find a PBH is one at small distances, since we start by assuming the existence of a PBH there.}

Instead of $p_c$, clustering is often given in terms of $\xi_\text{PBH}$, the excess contribution over the Poissonian one, see e.g. \cite{Animali:2024jiz,Auclair:2024jwj}.
This quantity is
also dubbed the two-point correlation function. In our case, it reads
\begin{equation} \label{eq:clustering_xi}
\begin{aligned}
    p_c(N_1,N_2;N_c) &= \beta(N_2;0,\theta_0)\times\qty[1 +\xi_\text{PBH}(N_1,N_2;N_c)] \\
    \Rightarrow \xi_\text{PBH}(N_1,N_2;N_c) &= \sqrt{\frac{N_1 N_2}{N_1 N_2 - N_c^2}}e^{\frac{(\pi-\theta_0)^2 N_c(2N_1 N_2 - N_2 N_c - N_2 N_c)}{2q^2 N_1 N_2 (N_1 N_2 - N_c^2)}} - 1 \, .
\end{aligned}
\end{equation}
Note that this only depends on the model parameters $q$ and $\theta_0$. Since $p_c$ has a more transparent physical interpretation than $\xi_\text{PBH}$, we use the former quantity below.

\begin{figure}
    \centering
    \includegraphics{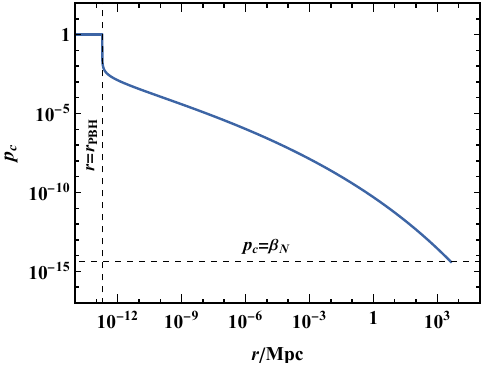}
    \caption{\it Clustering of PBHs for point A from Table~\ref{tab:benchmarks}, in terms of the comoving distance. The lower cutoff $r_\text{PBH} = \SI{1.9e-13}{Mpc}$ corresponds to the PBH mass scale $M=\SI{e21}{g}$.}
    \label{fig:clustering}
\end{figure}

Figure~\ref{fig:clustering} depicts $p_c$ for our model A from Table~\ref{tab:benchmarks} for two PBHs of equal mass $M_1=M_2=\SI{e21}{g}$, as a function of the comoving distance $r$. Right above the PBH radius $r_1=r_2=r_\text{PBH}$, the function quickly decreases to $\sim 10^{-3}$, indicating relatively strong clustering. This scale is essentially set by the prefactor of Eq.~\eqref{eq:p_of_second_PBH} (the exponent is close to zero), dominated by $b \ll 1$. Moving further out, $p_c$ decays towards the background value of $\beta \sim 10^{-15}$. The decay is notably mild: even though the absolute value of $p_c$ is small, it stays well above the background value over a wide range of scales.

Starting 
from $p_c$, we can compute the expected density of PBHs of mass $M_2$ at distance $r$ from the first PBH, assuming the additional PBHs are not correlated with each other. The number density in a logarithmic $r$ interval is
\begin{equation} \label{eq:n_PBH}
\begin{aligned}
    n_\text{PBH}(r) &= \frac{1}{\dd \ln r}\frac{\text{[comoving volume between $r$ and $r + \dd \ln r$]}}{\text{[comoving size of one PBH]}} \times p_c(N_1,N_2;N_c(r)) \\
    &= 3\qty(\frac{r}{r_2})^3\times p_c(N_1,N_2;N_c(r)) \,
\end{aligned}
\end{equation}
for $r > \max(r_1,r_2)$, and zero for $r \leq \max(r_1,r_2)$.
If the PBH density is small, as long as $\int_{0}^r \dd \ln r' \, n_\text{PBH}(r') \ll 1$, this is practically equal to the probability density for the distance to the closest PBH neighbour of mass $M_2$. The distance at which $\int_0^r \dd \ln r' \, n_\text{PBH}(r') = 1$ gives, roughly, the expected distance to this nearest neighbour. In practice, the integrand changes quickly with the geometric $r^3$ factor; the nearest-neighbour radius is reached when the integrand itself is 
of unit value. 
For our example model from Figure~\ref{fig:clustering}, $M_1=M_2=\SI{e21}{g}$, this is reached very close to the PBH, around $r \approx \SI{2e-12}{Mpc}$, only one order of magnitude larger than the PBH radius. The radius is essentially given by---up to an order one factor---by $r \sim r_2 \times [b/(2\pi |\Cc_{l,\text{th}}|)]^{-1/3}$. In comparison, the Poissonian limit $p_c=\beta_{N_2}$ gives $r \approx \SI{e-8}{Mpc} \sim r_2 \times \beta_{N_2}^{-1/3}$, five orders of magnitude larger than the PBH radius.

In fact, the PBH density in the clusters is so high that it quickly grows to dominate the local Universe.
The local PBH fraction within radius $r$ can be estimated as $p_c(r) \times {a}/{a_\text{formation}}$, where the redshift factor takes into account the relative dilution of PBHs versus radiation. Evaluated at Hubble re-entry, this becomes $p_c(r) \times {r}/{r_\text{PBH}}$, a growing function of $r$ that exceeds unity at $r \approx \SI{2e-8}{Mpc}$. On larger scales, the local Universe becomes matter-dominated before the scale's Hubble re-entry. The evolution of the thus-formed matter-dominated clusters is highly non-linear, and it can affect e.g. PBH mergers and subsequent PBH formation in non-trivial ways that go beyond the standard analysis \cite{Young:2019gfc, Raidal:2024bmm}. Analysing the  evolution
of PBH clusters is beyond the scope of this paper.

Fig.~\ref{fig:clustering} indicates that strong clustering extends all the way to the CMB pivot scale $r_*= \SI{20}{Mpc}$ and beyond. This poses a problem: clustered PBHs constitute isocurvature perturbations with a power spectrum roughly given by $\Pp_S(k) \sim \xi_\text{PBH}(r=1/k)$, giving $\Pp_S(k_*) \sim \xi_\text{PBH}(r_*) \approx 600$ in our model, breaking the observational bound \eqref{eq:P_S_bound} if PBHs constitute all dark matter. The probability of finding another PBH at distance $r_*$ is of order $600$, far exceeding the observed small CMB perturbations. In fact, we expect such behaviour to be generic to PBH models with a flat power spectrum like ours, including those considered in \cite{Stamou:2023vft, Stamou:2023vwz, Stamou:2024xkk}: all scales contribute equally to the final field perturbation that produces the PBHs, and thus clustering is strong over all scales, with a two-point function that grows in a power-law fashion as distance decreases.\footnote{In models with a peaked power spectrum, the scales outside the peak contribute less, presumably limiting the clustering range.} To be compatible with observations, the stochastic evolution of the curvaton $\psi$ must start later during inflation, when the CMB scales have exited the horizon, to decouple the CMB from the PBH statistics. This can be achieved, for 
example, by adding an inflaton-dependent mass to the curvaton field, large at early times to keep the curvaton stationary, but becoming smaller later, to allow for the curvaton's stochastic evolution. In the formulas of this paper, this corresponds to shifting the origin of the e-folds $N$ to the later, post-CMB time, and imposing a cutoff there. While affecting the details, such a shift does not alter the qualitative conclusions of this paper. Even with these measures, the PBHs may induce sizeable isocurvature perturbations at scales below the CMB, possibly altering the late-time formation of PBHs and leading to GW production. Again, such complications are beyond the scope of this paper.



If the underlying perturbations were Gaussian, the PBH distribution would be Poissonian, with $\xi_\text{PBH}(r) = 0$ at $r>r_\text{PBH}$ \cite{Ali-Haimoud:2018dau}. Extra clustering induced by non-Gaussianities has been studied before in e.g. \cite{Tada:2015noa, Franciolini:2018vbk, Desjacques:2018wuu, Suyama:2019cst}\footnote{Such can leave its imprints on scalar-induuced Gravitational Waves as shown in Refs.\cite{Papanikolaou:2024kjb,He:2024luf}.}, in a perturbative setup characterized by the parameter $f_{\rm NL}$ and its higher-order counterparts. These studies have shown that stronger non-Gaussianity implies a stronger coupling between the different scales and, thus, higher clustering. Our setup goes beyond a perturbative expansion and computes $\xi_\text{PBH}(r)$ from first principles. As is clear from the above discussion, the clustering in our model is stronger than in typical setups considered in the literature. Nevertheless, our conclusions concerning non-Gaussianity agree with the literature: In Appendix~\ref{sec:curvature_distribution}, we estimate the perturbative parameter $f_{\rm NL,\text{tot}}$ in our model, showing that it grows for large values of the parameters $b$ and $q$. By Eq.~\eqref{eq:p_of_second_PBH}, large $b$ and $q$ imply a high value of $p_c$ that decreases slowly as a function of distance. Thus, high perturbative non-Gaussianity is correlated with strong clustering.


Before us, clustering with non-perturbative non-Gaussianity was studied in \cite{Animali:2024jiz}\footnote{See also \cite{Shinohara:2021psq}, where the authors consider non-Gaussian clustering of supermassive PBHs in a simple threshold model.}, where the authors considered single-field models of inflation and used a PBH formation criterion based on a threshold for the curvature perturbation. They found the exponential tails of the curvature distribution to factorize the clustering profile $\xi_\text{PBH}$ so that it does not depend on the collapse threshold. Our model also produces an exponential tail for the curvature perturbation, see Appendix~\ref{sec:curvature_distribution}, and while we study PBH formation in terms of the compaction function instead of the curvature amplitude, our result \eqref{eq:clustering_xi} again does not depend on our threshold $\Cc_{l,\text{th}}$. This strengthens the hypothesis of a universal clustering profile in the presence of non-Gaussianity.

\medskip

\medskip

\section{Discussion and Conclusions}
\label{sec:discussion_and_conclusions}
In this work, we investigated a mechanism of primordial black hole production that does not
require a large amplitude for the curvature power spectrum, extending  ideas proposed in \cite{Carr:2019hud, Stamou:2023vft, Stamou:2023vwz, Stamou:2024xkk}.  
In Section \ref{sec_genfor}, we considered  the early Universe dynamics of a light spectator field, curvaton, whose energy density remains subdominant  during and after inflation. Rare field fluctuations probe a maximum of the spectator potential, leading to non-Gaussian tails in the induced curvature fluctuations. Such tails are  responsible for PBH formation, a process best studied in terms of the statistics of the fluctuations' compaction function ${\cal C}$. In Sections \ref{sec_axionfields} and \ref{sec:analytical_solutions},  we applied our formalism to a specific realisation based on the dynamics of axion-like particles (ALPs), analytically solving the evolution equations and deriving analytical formulas for the compaction function distribution \eqref{eq:Ccl_distribution_tail} and the PBH abundance \eqref{eq:beta_integrated}. In Appendix~\ref{sec:general_hilltop_computation}, we generalized the computation to other hilltop potentials.
This framework requires reduced fine-tuning of the parameters to produce PBHs compared to more conventional scenarios of single-field inflation.

In Section \ref{sec:benchmark_points}, we studied in detail three benchmark points, see Table~\ref{tab:benchmarks} and Fig.~\ref{fig:mass_distributions}, showing in the first two cases that they lead to PBH populations in the asteroid-mass window $10^{17} \dots 10^{22}$ g comprising all of the dark matter, while the third case leads to a subdominant PBH dark matter component of larger masses $\sim 10^{29}$ g, testable by upcoming experiments. Interestingly, we find a correlation between the ALP mass and decay constant, versus the PBH mass. For instance, ALP with mass $m_a = \SI{5.4e14}{eV}$ and decay constant $f_a = \num{4.6e-5}\Mpl$ leads to PBHs of mass $M_\text{PBH} = \SI{e21}{g}$ as the entire DM candidate of the universe, while $m_a = \SI{1.2e8}{eV}$ and decay constant $f_a = \num{3.8e-6}\Mpl$ leads to PBHs of mass $M_\text{PBH} = \SI{e29}{g}$, testable in future PBH observations via lensing in NGRST telescope and mergers detectable in Gravitational Waves (GW) detectors like LISA and ET (see~Fig.\ref{fig:mass_distributions}). 
In Section \ref{sec:applications}, finally, we studied further developments of our  set-up. We  considered mixed ALP and PBH dark matter scenarios, the dynamics of spectators with Higgs-like potentials, and the phenomenon of PBH clustering after formation. The latter process turns out to be particularly important for the phenomenology of our set-up: we find that PBHs cluster strongly over all cosmological scales, clashing with CMB isocurvature bounds. This problem is shared by all inflationary PBH models that depend  on strongly non-Gaussian cosmological fluctations, without a peak in the curvature power spectrum. We then outlined possible improvements
in  model building for preventing  this issue.
 
There are a number of things to keep in mind regarding our analysis. In particular, our analytical approximations are valid for sharp features in the profile of the curvature perturbation, and it is not clear whether such features would appear in a more comprehensive analysis. Instead, perturbations we classify as black holes may lead to patches of space where the curvaton is pushed off the potential maximum to its other side. Such patches may still form PBHs or other bound structures, especially if the field gets stuck in a false vacuum on the other side of the maximum. They may also produce topological defects, which later lead to PBH production \cite{Belotsky:2018wph, Gelmini:2022nim,Ge:2023rrq}. In addition, our analysis is uncertain for PBHs that form after the field has started to oscillate but before it decays. A detailed analysis of this isocurvature regime is beyond the scope of this work.

Besides further theoretical  developments needed in  model building, there are several  avenues for further investigations. The correlations between ALP properties and the mass of the PBHs suggest a synergy between cosmological and astronomical searches for PBHs and laboratory experiments aimed to ALP detection. When it comes to gravitational waves, our set-up does not generate a sizeable amount of induced gravitational waves at second order in perturbations, since the curvature power spectrum has a small amplitude at all scales. Nevertheless, given the strong PBH clustering effects we find, we  expect significant GW signals from PBH mergers, whose properties warrant further investigation. 


\bigskip

\section*{Acknowledgements}

C.C. is supported by NSFC (Grants No. 12433002).
The work of G.T. is partially funded by STFC grant ST/X000648/1. This work was supported by the Estonian Research Council grant PRG1055 and by the EU through the European Regional Development Fund CoE program TK133 ``The Dark Side of the Universe.'' E.T. was supported by the Lancaster--Manchester--Sheffield Consortium for Fundamental Physics under STFC grant: ST/T001038/1. For the purpose of open access, the authors have applied a Creative Commons Attribution licence to any Author Accepted Manuscript version arising. Research Data Access Statement: No new data were generated for this manuscript. Authors thank Devanshu Sharma and Spyros Sypsas for discussion. C.C. thanks the support from the Jockey Club Institute for Advanced Study at The Hong Kong University of Science and Technology.

\begin{appendix}

\section{Compaction function distribution at small $\Cc_l$}
\label{sec:small_compaction}
How does the compaction function probability distribution $P(\Cc_l,N)$ of Eq.~\eqref{eq:Ccl_distribution_scaled} behave at small $\Cc_l$ values? To approach this question, let us first compute the variance of $\Cc_l$:
\begin{equation} \label{eq:Ccl_variance}
    \expval{\Cc_l^2} = \int_{-\infty}^\infty \dd \Cc_l \, \Cc_l^2 \, P(\Cc_l,N)
    = \frac{q}{\sqrt{2\pi N}}\int_{-\infty}^{\infty} \dd \ttheta \tilde{N}_{\ttheta}^2 \exp[-\frac{(\ttheta-\theta_0)^2}{2q^2N}] \approx q^2\tilde{N}_{\theta_0}^2 = \Pp_{\zeta,\psi} \, .
\end{equation}
The approximation follows from the fact that the Gaussian factor only has support around $\ttheta = \theta_0$. It is accurate if $\tilde{N}_{\ttheta}$ behaves regularly enough. However, around $\ttheta \sim \pi$, $\tilde{N} \sim \ln(\pi-\ttheta)$ (and similarly around other potential maxima), see Eq.~\eqref{eq:Ntilde_scaling}, and the integral actually diverges. Result \eqref{eq:Ccl_variance} still describes the distribution at its bulk, at not-too-large $\Cc_l$, if the asymptotic tails are disregarded. Since the observable Universe is finite, we expect cosmological observations not to probe the tails far enough to affect $\expval{\Cc_l^2}$, and Eq.~\eqref{eq:Ccl_variance} applies. It is essentially a linear perturbation theory result, as suggested by the appearance of the power spectrum $\Pp_{\zeta,\psi}$.

A reasonable guess for $P(\Cc_l,N)$ at small $\Cc_l$ is a Gaussian with a variance given by Eq.~\eqref{eq:Ccl_variance}:
\begin{equation} \label{eq:Ccl_distribution_Gaussian}
    P(\Cc_l,N) = \frac{1}{\sqrt{2\pi} q \tilde{N}_{\theta_0}} \exp( -\frac{\Cc_l^2}{2q^2\tilde{N}^2_{\theta_0}})
    = \frac{1}{\sqrt{2\pi \Pp_{\zeta,\psi}}} \exp( -\frac{\Cc_l^2}{2\Pp_{\zeta,\psi}}) \, .
\end{equation}
This can be obtained from Eq.~\eqref{eq:Ccl_distribution_scaled} by assuming the first term dominates in the exponent and integrating over it---the opposite of how we computed the tail in Eq.~\eqref{eq:Ccl_distribution_tail}.

\begin{figure}
    \centering
    \includegraphics[scale=1]{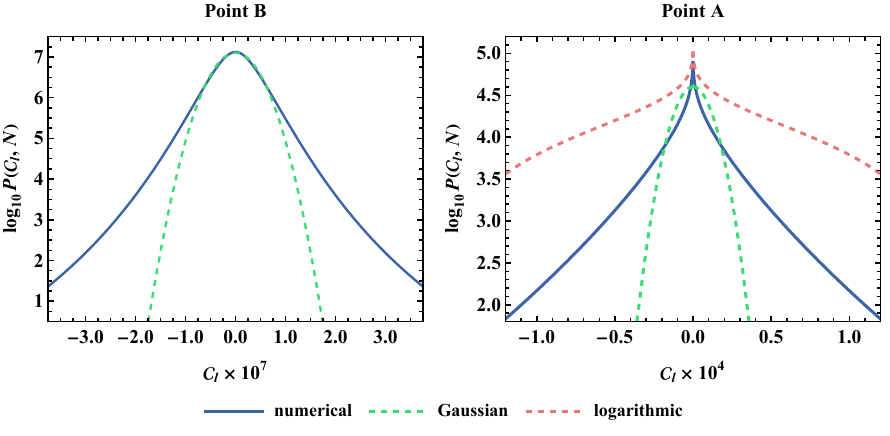
    }
    \caption{\it The probability distribution $P(\Cc_l,N)$ near $\Cc_l=0$ for the benchamrk points A and B from Table~\ref{tab:benchmarks}. The Gaussian approximation refers to Eq.~\eqref{eq:Ccl_distribution_Gaussian}; the logarithmic approximation refers to Eq.~\eqref{eq:Ccl_distribution_small_theta}.}
    \label{fig:Clmid}
\end{figure}

Fig.~\ref{fig:Clmid} plots the approximation \eqref{eq:Ccl_distribution_Gaussian} together with a numerically integrated $P(\Cc_l,N)$ for two of our benchmark points. We see that for point B, the match is reasonable: there is visible non-Gaussianity even near the peak, but a general shape with the expected variance appears. However, for point A, the fit is not good; instead, $P(\Cc_l,N)$ appears to diverge for small $\Cc_l$. This is due to the small $\theta_0$ value: for point A, $\theta_0 = 0.2$, and $q\sqrt{N} \approx 0.4$ is large enough for the distribution to probe the $\ttheta=0$ region, where $\tilde{N}_{\ttheta}$ approaches zero and the $1/\tilde{N}_{\ttheta}$ factor in Eq.~\eqref{eq:Ccl_distribution_scaled} grows. The first term in the exponent of Eq.~\eqref{eq:Ccl_distribution_scaled} is not sharp enough to dominate the integrand for small $\Cc_l$; instead, the $\tilde{N}_{\ttheta}$ factors start to play a role.

We can try to tackle the problem like we did for large $\Cc_l$ values in Section~\ref{sec:C_distribution}, with the saddle point approximation \eqref{eq:saddle_point_approx}. Indeed, the $\tilde{N}_{\ttheta}$ factors form a peak at small $\ttheta$. However, the integrand is much wider in this region than near the hilltop, with a $1/\tilde{N}_{\ttheta} \sim 1/\ttheta$ tail, see Eq.~\eqref{eq:decay_N_eq_small_theta}, so the Gaussian approximation of Section~\ref{sec:C_distribution} doesn't work well. Instead, we can estimate the $\exp[-\Cc_l^2/(2q^2\tilde{N}_{\ttheta}^2)]$ factor as a step function that kills the integrand at $|\ttheta| < \ttheta_\text{pk}$ and is one elsewhere. Using $\tilde{N}$ from Eq.~\eqref{eq:decay_N_eq_small_theta} and taking $\ttheta -\theta_0 \approx -\theta_0$ in the first term in the exponent, we get
\begin{equation}
\begin{aligned}
\label{eq:Ccl_distribution_small_theta}
    P(\Cc_l,N) &= \frac{4 R_\text{dec}^{1/4}}{\pi q^2\sqrt{N}f_a^{3/2}}\exp(-3n_\text{osc}-\frac{\theta_0^2}{2q^2 N})\int_{\ttheta_\text{pk}}^\pi \frac{\dd \ttheta}{\ttheta} \\
    &= \frac{4 R_\text{dec}^{1/4}}{\pi q^2\sqrt{N}f_a^{3/2}}\exp(-3n_\text{osc}-\frac{\theta_0^2}{2q^2 N}) \ln \frac{\pi f_a^{3/2}q e^{3n_\text{osc}}}{4|\Cc_l| R_\text{dec}^{1/4}} \, ,
\end{aligned}
\end{equation}
where we used the symmetry in $\ttheta \leftrightarrow -\ttheta$. As in Section~\ref{sec:C_distribution}, we also restricted the integration to $|\ttheta| < \pi$.
The result diverges at $\Cc_l \to 0$, but the variance $\expval{\Cc_l^2}$ is still finite. The approximation \eqref{eq:Ccl_distribution_small_theta} is compared to the numerically integrated result in Fig.~\ref{fig:Clmid}. The match is not perfect, but the qualitative behaviour matches: instead of a wide hilltop, we obtain a sharp, diverging peak.

\section{PBHs from a general hilltop curvaton}
\label{sec:general_hilltop_computation}
In Section~\ref{sec:analytical_solutions}, we solved the initial PBH fraction $\beta$ in the axion-like curvaton model. An analogous computation applies to any curvaton $\psi$ with a potential $V$ that has a quadratic minumum of zero at some $\psi_\text{bottom}$ and a quadratic maximum at some $\psi_\text{top}$. In this appendix, we generalize the results of Section~\ref{sec:analytical_solutions} to such a model. Without loss of generality, we take $\phi_\text{bottom} < \phi < \phi_\text{top}$. 

Let us start by fixing the notation and rescaling the variables, analogously to Eq.~\eqref{eq:scaled_variables} :
\begin{equation} \label{eq:general_rescalings}
\begin{gathered}
    \Delta\psi \equiv \psi_\text{top}-\psi_\text{bottom} \, , \quad
    \phi \equiv \frac{\psi}{\Delta\psi} \, , \quad
    f(\phi) \equiv \frac{V(\psi)}{V_\text{top}} \, , \quad
    \tilde{R} \equiv \frac{\rho_r}{V_\text{top}} \, , \\
    Q \equiv \frac{H_*}{2\pi\Delta\psi} \, , \quad  
    V_\text{top} \equiv V(\psi_\text{top}) \, , \quad
    \mu^2 \equiv -V''(\psi_\text{top}) \, , \quad
    m^2 \equiv V''(\psi_\text{bottom}) \, .
\end{gathered}
\end{equation}
For convenience, some of these scalings differ from the ALP case slightly; we comment this further below. In general, the indices ``top'' and ``bottom'' refer to the hilltop and mimum of the potential, respectively.

Analogously to Eqs.~\eqref{eq:theta_eom} and \eqref{eq:power_spectrum_theta},
the equation of motion for the scaled field $\phi$ is
\begin{equation} \label{eq:general_phi_eom}
    \phi'' + \qty(3-\frac{1}{2}\phi'^2 \Delta\psi^2)\qty(\qty[1-\frac{2}{3}\frac{\tilde{R}}{\tilde{R}+f(\phi)}]\phi' + \frac{f'(\phi)}{\qty[\tilde{R}+f(\phi)]\Delta\psi^2}) = 0 \, , \quad
    \tilde{R} = \tilde{R}_\text{dec} e^{-4N_p} \, ,
\end{equation}
and its power spectrum contribution is
\begin{equation} \label{eq:general_power_spectrum_phi}
    \Pp_{\zeta,\psi}(k) = Q^2 \tilde{N}_{\tilde{\phi}}^2(\phi_0) \, , \quad \phi_0 \equiv \frac{\psi_0}{\Delta\psi} \, ,
\end{equation}
where $\tilde{N}(\tilde{\phi})$ is the time to the curvaton decay surface starting from $\tilde{\phi}$, as above. When $\tilde{\phi}$ is near the hilltop, we can again define the new time variable $n \equiv \frac{1}{4}\ln\frac{\mu^2}{\tilde{R}(N)V_\text{top}}$, so that Eq.~\eqref{eq:general_phi_eom} becomes, approximately,
\begin{equation} \label{eq:general_phi_approx_eom}
    \phi'' + \phi' - 3(\phi-\phi_\text{top}) e^{4n} = 0
\end{equation}
with the solution
\begin{equation} \label{eq:general_phi_solution}
\begin{aligned}
    \phi_\text{top} - \phi(n) &= (\phi_\text{top} - \tilde{\phi}) \times \frac{\sqrt{2}}{3^{1/8}}\Gamma\qty(\frac{5}{4}) e^{-\frac{n}{2}} I_{1/4}\qty(\frac{\sqrt{3}}{2}e^{2n}) \, .
\end{aligned}
\end{equation}
We approximate the oscillations to start when $\psi = \psi_\text{bottom}$, that is, $\phi_\text{top} - \phi(n) = 1$. This gives
\begin{equation} \label{eq:general_n_osc}
    n_\text{osc} \approx
    -\frac{2}{3}\ln(\frac{\sqrt{\pi} 3^{3/8}}{\sqrt{2}(\phi_\text{top}-\tilde{\phi})\Gamma\qty(\frac{5}{4})})
    -\frac{1}{2}W_{-1}\qty(-\frac{2^{5/3}\qty[(\phi_\text{top}-\tilde{\phi})\Gamma\qty(\frac{5}{4})]^{4/3}}{3\pi^{2/3}}) \, ,
\end{equation}
analogously to Eq.~\eqref{eq:n_osc_in_phi_ini}. Similarly to Eq.~\eqref{eq:Ntilde_approx_3}, noting that $f=1$ on the hilltop, we can use this to write
\begin{equation} \label{eq:general_Ntilde_approx_3}
    \tilde{N} \approx \frac{1}{4\tilde{R}^{1/4}_\text{dec}}\qty(\frac{V_\text{top}}{\mu^2})^{3/4}e^{3n_\text{osc}} \, .
\end{equation}
We can again approximate
\begin{equation} \label{eq:general_Ntilde_scaling}
    \tilde{N}(\phi) = a - b\ln (\phi_\text{top} - \tilde{\phi}) \, ,
\end{equation}
now with
\begin{equation} \label{eq:general_b}
    b = (\phi_\text{top} - \tilde{\phi})\tilde{N}_{\tilde{\phi}} = \frac{3^{1/4}V^{3/4}_\text{top}}{4\sqrt{2} \mu^{3/2} \tilde{R}_\text{dec}^{1/4}} \times \sqrt{-W_{-1}\qty(-\frac{3 
q^4V^3_\text{top}\Gamma\qty(\frac{5}{4})^4}{32|\Cc_{l,\text{th}}|^4 \mu^6 \pi^2 R_\text{dec}})}
    \quad
    \text{for} \quad \tilde{N}_{\tilde{\phi}} = \frac{|\Cc_{l,\text{th}}|}{q} \, .
\end{equation}
Analogously to Eq.~\eqref{eq:beta_integrated}, this helps us integrate over the tail of the $\Cc_l$ distribution, yielding
\begin{equation} \label{eq:general_beta_integrated}
    \beta_N \approx \frac{b}{2\pi\sqrt{N}|\Cc_{l,\text{th}}|} \exp[-\frac{(1-\phi_0)^2}{2q^2N}] \, .
\end{equation}

To reconnect with the ALP model, we identify
\begin{equation} \label{eq:axion_setup}
\begin{gathered}
    \Delta\psi = \pi f_a \, , \quad
    \phi = \frac{\theta}{\pi} \, , \quad
    f(\phi) = 1-\cos(\pi\phi) \, , \quad
    Q = \frac{H_*}{2\pi^2 f_a} = \frac{q}{\pi} \, , \\
    \tilde{R}_\text{dec} = \frac{1}{2}R_\text{dec} \, , \quad
    V_\text{top} = 2\Lambda^4 \, , \quad
    m^2 = \mu^2 = m_a^2 = \frac{\pi^2}{2}\frac{V_\text{top}}{\Delta\psi^2} \, .
\end{gathered}
\end{equation}
Note, in particular, the slight differences in $\Delta \psi$ and $f_a$, $Q$ and $q$, and $\tilde{R}_\text{dec}$ and $R_\text{dec}$ (arising from comparing the energy densities to $V_\text{top}$ instead of $\Lambda_a^4 = V_\text{top}/2$). Importantly, in the ALP model, the parameters $\Delta\Psi$, $V_\text{top}$, and $m_a$ are linked together so that only two of them can be set freely.

For the Higgs-like field of Section \ref{sec:higgs_PBHs}, we have
\begin{equation} \label{eq:higgs_setup}
\begin{gathered}
    \Delta\psi = v \, , \quad f(\phi) = (\phi^2-1)^2 \, , \quad Q = \frac{H_*}{2\pi v} \, , \\
    V_\text{top} = \frac{\lambda}{4}v^4 \, , \quad m^2 = 2\lambda v^2 = 8\frac{V_\text{top}}{\Delta\psi^2} \, , \quad \mu^2 = \lambda v^2 = \frac{1}{2}m^2 = 4\frac{V_\text{top}}{\Delta\psi^2} \, .
\end{gathered}
\end{equation}
The only difference between the models is the slight difference in the shape of the $f$ function, reflected in the different (order one) coefficients between $V_\text{top}/\Delta\psi^2$, $m^2$, and $\mu^2$, which are still related to each other. The PBH behaviour is essentially the same.

In general, PBH abundance is still mostly regulated by $Q$, as above, and $b$ still has to be small to make $\Pp_{\zeta,\psi}$ small. If we want $b$ to be as large as possible without violating the $\Pp_{\zeta,\psi}$ bound, we had to go to small $R_\text{dec}$ above, since our $f_a \sim q H_*$ was capped by the maximum allowed value of $H_*$. This limitation still applies for our $\Delta\psi$, but now $b$ is instead related to the ratio $V_\text{top}/\mu^2$, which can, at least in principle, be independent of $\Delta \psi$. In particular, making $\mu$ small increases $b$ without the need to push $R_\text{dec}$ down. This corresponds to a very wide  hilltop, necessitating a sharp drop towards the minimum at distance $\Delta \psi$, suggesting a plateau-type model.

\medskip

\section{Probability distribution of the curvature perturbation}
\label{sec:curvature_distribution}
As discussed around Eq.~\eqref{eq:delta_N_formula}, the curvature perturbation is given by the $\delta\mathcal{N}$ formalism as
\begin{equation} \label{eq:zeta_delta_N}
    \zeta = \Delta N = \tilde{N}(\ttheta) - \tilde{N}(\theta_0) \, ,
\end{equation}
in terms of the field value $\ttheta$. The field's probability distribution at $N$ inflationary e-folds, $P(\ttheta,N)$, is given by Eq.~\eqref{eq:theta_stochastics}.
From this, we get a probability distribution for $\zeta$:
\begin{equation} \label{eq:P_Delta_N_from_P_theta}
    P(\Delta N, N) = \frac{P(\ttheta(\Delta N), N)}{\Delta N_{\ttheta}(\ttheta(\Delta N))} \, ,
\end{equation}
where the functional dependence between $\Delta N$ and $\ttheta$ arises from Eq.~\eqref{eq:zeta_delta_N}.

In the tail of the distribution, near $\ttheta \to \pi$ and large $\Delta N$, we can use Eq.~\eqref{eq:Ntilde_scaling}, redefining the coefficient $a$:
\begin{equation} \label{eq:Delta_N_in_theta_tail}
    \Delta N = \Delta N_1 - b\ln\frac{\pi-\ttheta}{\pi-\theta_1} \, ,
\end{equation}
where $\Delta N_1 \equiv \Delta N(\theta_1)$ must be fixed numerically. In this Section, we don't care about its exact value; we're only interested in the qualitative asymptotic behaviour of Eq.~\eqref{eq:P_Delta_N_from_P_theta}, which is now
\begin{equation} \label{eq:P_Delta_N_tail}
    P_\text{tail}(\Delta N, N) \approx \frac{\pi-\theta_1}{\sqrt{2\pi N} q b}\exp(-\frac{(\pi-\theta_0)^2}{2q^2N} + \frac{\Delta N_1 - \Delta N}{b}) \, .
\end{equation}
Note the exponential tail $\exp(-\Delta N/b)$, which gives extra exponential suppression compared to the distribution of $\Cc_l$, Eq.~\eqref{eq:Ccl_distribution_tail}. This comes, essentially, from the computation $1/\tilde{N}_{\ttheta} \propto \pi - \ttheta \propto e^{-\tilde{N}/b}$, compared to the computation $1/\tilde{N}_{\ttheta\ttheta} \propto (\pi - \ttheta)^2 \propto 1/\tilde{N}_{\ttheta} \propto 1/\Cc_l$. $\Delta N$ behaves like $\tilde{N}$, whereas our $\Cc_l$ behaves like $\tilde{N}_{\ttheta}$. In particular for small $b$, the $\Delta N$ distribution is much suppressed compared to the $\Cc_l$ distribution. In this limit, one has to go closer to the hilltop at $\ttheta=\pi$ to get a high $\Delta N$ than to get a high $|\Cc_l|$. Physically, for a large $\Cc$, we don't need a large number of e-folds, we only need a momentary rapid change in the e-folds, which can take place over a short distance. This is also evident in our benchmark points in Table~\ref{tab:benchmarks}, and partially explains the high PBH abundances we obtain: we look at $\Cc_l$ instead of the harder-to-increase $\Delta N$, like many previous studies do.

Similar exponential tails arise also from single-field models of stochastic inflation \cite{Pattison:2017mbe, Ezquiaga:2019ftu, Biagetti:2021eep}, including potentials with a local maximum in the potential \cite{Figueroa:2020jkf, Figueroa:2021zah, Tomberg:2022mkt,Tomberg:2023kli, Briaud:2023eae}. Also in these models, considering $\Cc_l$ instead of $\Delta N$ seems to amplify the probability distribution \cite{Raatikainen:2023bzk}.

If, instead of the tail, we look at the typical values $\ttheta \approx \theta_0$, small $\Delta N$, we can expand
\begin{equation} \label{eq:Delta_N_expanded}
    \Delta N \approx \tilde{N}_{\ttheta}(\theta_0)(\ttheta - \theta_0) + \frac{1}{2}\tilde{N}_{\ttheta\ttheta}(\theta_0)\qty[(\ttheta-\theta_0)^2 - \expval{(\ttheta-\theta_0)^2}] + \mathcal{O}[(\ttheta-\theta_0)^3] \, .
\end{equation}
Defining $\Delta N_G \equiv \tilde{N}_{\ttheta}(\theta_0)(\ttheta - \theta_0)$, which is Gaussian, we can rearrange this into
\begin{equation} \label{eq:Delta_N_fNL}
    \Delta N = \Delta N_G + \frac{3}{5}f_{NL}\qty(\Delta N_G^2 - \expval{\Delta N_G^2}) \, , \qquad f_{NL} = \frac{5\tilde{N}_{\ttheta\ttheta}(\theta_0)}{6\tilde{N}_{\ttheta}^2(\theta_0)} \, .
\end{equation}
Here $f_{NL}$ is the usual bispectrum parameter of local non-Gaussianity \footnote{Refs. \cite{Palma:2019lpt,Celoria:2021vjw,Ferrante:2022mui,Creminelli:2024cge} finds that $f_nl$, $g_nl$ expansion breaks down far in the  tail. } - - it is scale invariant and would be the $f_{NL}$ parameter measured from, say, the CMB, if the scalar field perturbations dominated. Table~\ref{tab:fNL} gives the values for our benchmark points, all of them large. Indeed, Eqs.~\eqref{eq:Delta_N_fNL} with \eqref{eq:Ntilde_approx_3} suggests $f_{NL} \sim 1/\tilde{N}(\theta_0) \sim R^{1/4}_\text{dec}/f_a^{3/2} \gg 1$; using also Eq.~\eqref{eq:power_spectrum_theta} suggests $f_{NL} \sim 1/\tilde{N}_{\ttheta}(\theta_0) \sim q/\sqrt{\Pp_{\zeta,\psi}} \gg 1$.

\begin{table}
\begin{center}
\begin{tblr}{colspec={
    Q[1.8cm,halign=r]
    Q[1.8cm,halign=c]
    Q[1.8cm,halign=c]
    Q[1.8cm,halign=c]
  }}
\toprule
& A & B & C \\
\midrule
$f_{NL}$ & \num{2.8e4} & \num{1.3e6} & \num{1.5e5} \\
\bottomrule
\end{tblr}
\end{center}
\caption{\it The non-Gaussianity parameter $f_{NL}$ from Eq.~\eqref{eq:Delta_N_fNL} for the benchmark points of Table~\ref{tab:benchmarks}.}
\label{tab:fNL}
\end{table}

As a consistency check, we can compute the $f_{NL}$ in the quadratic potential limit of small $\theta$, with $\tilde{N}$ from Eq.~\eqref{eq:Ntilde_squared_small_theta}. We get
\begin{equation} \label{eq:fNL_check}
\begin{gathered}
    f_{NL} = \frac{5\times 4R_\text{dec}^{1/4}}{6\theta_0^2f_a^{3/2}e^{3n_\text{osc}}} = \frac{5}{4r} \, , \\
    r \equiv \frac{3\theta_0^2f_a^{3/2}e^{3n_\text{osc}}}{8R_\text{dec}^{1/4}} = \frac{3}{4}\frac{\frac{1}{2}\theta_0^2 \times \Lambda_a^4}{\Lambda_a^4e^{-4n_\text{osc}}/f_a^2}\qty(\frac{\Lambda_a^4e^{-4n_\text{osc}}/f_a^2}{\Lambda_a^4 R_\text{dec}})^{1/4} = \frac{3\rho_{\psi,\text{dec}}}{4\rho_{r,\text{dec}}} \, ,
\end{gathered}
\end{equation}
where $\rho_{\psi,\text{dec}}$ and $\rho_{r,\text{dec}}$ are the scalar and radiation densities at the scalar decay. We recognized $\frac{1}{2}\theta_0^2\Lambda_a^4$ and $\Lambda_a^4e^{-4n_\text{osc}}/f_a^2 = m_a^2e^{-4n_\text{osc}}$ as the energy densities of the scalar and radiation at the the start of the scalar oscillations, and the quantity in the parantheses is the redshift factor from that time to the decay, given by the radiation energy density. Result \eqref{eq:fNL_check} agrees with the well-known quadratic curvaton result \cite{Lyth:2002my}.

In our setup, the scalar field is subdominant when it decays, so the values listed in Table~\ref{tab:fNL} don't correspond to non-Gaussianity observed in, say, the CMB. The induced perturbative non-Gaussianity is suppressed and not of the local type. Still, we can compute a parameter like $f_{NL}$ by comparing the non-Gaussian term of Eq.~\eqref{eq:Delta_N_fNL} to the total curvature perturbation $\zeta$. Using the previous results, this yields $f_{NL,\text{tot}} \sim f_{NL,\psi} \times \Pp_{\zeta,\psi}/\Pp_{\zeta,\text{tot}} \sim q\sqrt{\Pp_{\zeta,\psi}}/\Pp_{\zeta,\text{tot}}$. Current CMB observations indicate $f_{NL,\psi} \lesssim \mathcal{O}(10)$ \cite{Planck:2019kim}; with $q\sim0.01$ and $\Pp_{\zeta,\text{tot}} \sim 10^{-9}$, this sets the requirement $\Pp_{\zeta,\psi} \lesssim 10^{-12}$. This is not satisfied for our benchmark point A. However, as discussed in Section \ref{sec:clustering}, couplings between the CMB and PBH scales lead to large PBH clustering, providing even more stringent constraints. In that Section, we discussed how to remove these couplings; the same measures also erase $f_{NL}$ at the CMB scales.

\end{appendix}

\bibliographystyle{JHEP}
\bibliography{ref}
\end{document}